\documentclass{aa}
\usepackage{afterpage}
\usepackage{amsmath}
\usepackage{booktabs}
\usepackage{dcolumn}
\usepackage{graphicx}
\usepackage{longtable}
\usepackage{lscape}
\usepackage{rotating}
\usepackage[tight]{subfigure}
\usepackage{supertabular}

\usepackage{graphicx}
\usepackage{natbib}
\usepackage{afterpage,amssymb,booktabs,dcolumn,longtable,lscape,palatino,rotating,subfigure,supertabular,txfonts}
\newcommand{\htwoo}{H$_2$O}
\newcommand{\methanol}{CH$_3$OH}
\newcommand{\Lsun}{L$_\odot$}

\newcommand{\nancay}{Nan\c{c}ay}
\newcommand{\s}{s$^{-1}$}

\begin{document}

\title{A Survey of OH Masers Towards High Mass Protostellar Objects}

\author{K.~A.~Edris\inst{1,2}
 \and G.~A.~Fuller\inst{1}
 \and R.~J.~Cohen\inst{3}}

\institute{The University of Manchester, School of Physics and
  Astronomy, Sackville Street Building, PO Box 88, Manchester, M60
  1QD, UK
\and 
Al-Azhar University, Faculty of Science, Astronomy
Department, PO Box 11884, Naser City, Cairo, Egypt
 \and The University of Manchester, Jodrell Bank Observatory, Macclesfield, Cheshire SK11 9DL, UK}

\authorrunning{Edris et al.}
\offprints{G.~A.~Fuller,
  \email{G.Fuller@manchester.ac.uk}}

\date{Received ; accepted } 

\abstract{Masers are important tracers of the early evolution of young
  high mass stars, but the relationship between different types of
  maser and the evolutionary state of the exciting source remains
  unclear.}{To determine whether OH masers are common towards
  candidate high mass protostellar objects.}{We present a survey of OH
  maser emission towards a sample of high mass protostellar objects
  made using the \nancay\ and GBT telescopes.}{ OH maser emission was
  detected towards 63 objects with 36 new detections. There are 56
  star-forming regions and 7 OH/IR candidates. Nearly half of the
  detected sources have OH flux densities $\lesssim$ 1~Jy.  There is
  no evidence that sources with OH masers have a different range of
  luminosities from the non-maser sources.  The results of this survey
  are compared with previous \htwoo\ and class II \methanol\ maser
  observations of the same objects.  Some of the detected sources are
  only associated with OH masers and some sources are only associated
  with the 1720~MHz OH maser line.  The velocity range of the maser
  emission suggests that the water maser sources may be divided into
  two groups. The detection rates and velocity range of the OH and {
    Class II} \methanol\ masers support the idea that there is a
  spatial association of the OH and { Class II} \methanol\ masers.
  The sources span a wide range in $R$, the ratio of the methanol
  maser peak flux to OH 1665~MHz maser peak flux, however there are
  only a few sources with intermediate values of $R$, $8<R<32$, which
  has characterised previous samples.  The majority of the sources are
  either methanol-favoured or OH-favoured.  Sources which have masers
  of any species, OH, water or methanol, have redder
  [100$\mu$m$-$12$\mu$m] IRAS colours than those without masers.
  However, there is no evidence for different maser species tracing
  different stages in the evolution of these young high mass
  sources.}{The detection of OH masers towards 26\% of a sample of 217
  sources should remove any doubt about the existence of OH maser
  emission towards these objects or this early evolutionary stage.
  Previous observations which have shown that the OH maser emission
  from similar sources traces the circumstellar disks around the
  objects. This combined with the sensitivity of the OH emission to
  the magnetic field, make the newly detected sources interesting
  candidates for future follow-up at high angular
  resolution.\thanks{The data shown in Figures~\ref{fig:spectra} and
    \ref{fig:spectra-offset} are available in electronic form at the
    CDS via anonymous ftp to cdsarc.u-strasbg.fr (130.79.128.5) or via
    http://cdsweb.u-strasbg.fr/cgi-bin/qcat?J/A+A/ }}

\keywords{masers -- stars: formation -- ISM: molecules --HII regions }
  
\titlerunning{OH Maser Survey Towards HMPOs}

\maketitle

\section{Introduction}

Compact HII regions, poorly collimated
bipolar molecular outflows and circumstellar disks are
signs of the existence of massive protostars (e.g. Garay \& Lizano
\citealp{garay99}; Churchwell \citealp{churchwell02}).  Maser
emission is also found to be associated with these objects (e.g.
Garay \& Lizano \citealp{garay99}; Edris et al. \citealp{Edris05})
with OH, \htwoo\ and \methanol\ the three most widespread types of
maser associated with these regions. These species have been used
as probes of star-forming regions as their maser emission provide
unique information on these dense dusty regions (e.g. Cohen
\citealp{cohen89}, and references therein).  Observations of
\htwoo\ and { Class II} \methanol\ masers have
shown that both maser types are signposts of high mass star
formation in very early evolutionary stages (Beuther et al.
\citealp{beuther02}; Szymczak et al. \citealp{SHK2000}, hereafter
SHK2000; Palla et al.  \citealp{palla91}, hereafter P91).

On the other hand OH masers are known to be associated with an
advanced stage of the appearance of UCHII region (e.g. Garay \& Lizano
\citealp{garay99}, and references therein). Models of these OH masers
assume that the maser arises in the compressed shell between the shock
and ionisation fronts around the HII region (Elitzur \& De~Jong
\citealp{elitzur78}).  However in a survey, Caswell \cite{caswell83}
found large proportion of OH masers have no closely related prominent
HII regions. In some molecular outflow sources, OH masers have been
mapped with high angular resolution and found to be associated with an
earlier stage of molecular outflows and circumstellar disks (Cohen,
Rowland \& Blair \citealp{cohen84}; Brebner \citealp{brebner88}; Cohen
et al.  \citealp{cohen03}; Edris et al. \citealp{Edris05}). Since the
masers can be observed with high spatial, and spectral, resolution
they can probe the inner regions of these sources. OH masers also
provide the possibility of measuring the magnetic fields in these
regions.

With this in mind, and to form a more complete picture of the
relationship between maser emission and the evolution of high mass
protostars, 217 high mass protostellar object (HMPO)
candidates have been surveyed for OH maser emission using the
\nancay\ radio telescope\footnote{The Nan\c{c}ay Radio Observatory
is the Unit\'e
  scientifique de Nan\c{c}ay of the Observatoire de Paris, associated
  as Unit\'e de Service et de Recherche (USR) No. 704 to the French
  Centre National de la Recherche Scientifique (CNRS).  The Nan\c{c}ay
  Observatory also gratefully acknowledges the financial support of
  the Conseil R\'egional de la R\'egion Centre in France.  } and the
NRAO Green Bank Telescope\footnote{The National Radio Astronomy
  Observatory (NRAO) is a facility of the National Science Foundation
  operated under cooperative agreement by Associated Universities,
  Inc.} (GBT).  The aims of this survey are:
\begin{itemize}
  \item to determine whether OH masers are associated with these
  objects,
  \item to investigate the relationship between the OH and H$_2$O and/or
  CH$_3$OH maser emission and whether the masers are related to the
  evolutionary stage of these objects or represent different
  regions of the star formation core,
\item to identify sources for further study at high angular
  resolution.
\end{itemize}
The description of the sample is given in section 2 and the details of
the observations are given in section 3. In section 4 we report the
results of the survey while section 5 presents some detection statistics.
In section 6 we discuss the interpretation while the conclusions are
drawn in section 7.

\section{The sample}

\begin{table*}
\begin{center}
\begin{tabular}{llllll}
 \toprule
 \toprule
Class of object &  [25$-$12]  &  [60$-$25] & [100$-$60] & [60$-$12] & Ref. \\
 \midrule
Cores           & 0.4$-$1.0   & 0.4$-$1.3    &  0.1$-$0.7    & $-$         & 1   \\
H$_{2}$O maser  & 0.5$-$1.1   & 0.4$-$1.7    & $-$0.1$-$0.5    & $-$         & 2   \\
UCHII regions   & $>$~0.6    &   $-$        &  $-$          & $>$~1.3    & 3   \\
Bright IRAS     & 0.5$-$1.2   & 0.6$-$1.6    &  0.0$-$0.6    &   $-$        & 4   \\
CMC+UCHII       & $\geq$~0.57& 0.61$-$1.74  &  0.087$-$0.52 & $\geq$~1.3 & 5   \\
OH maser        & $>$~1.2 &    $-$      &  $-$             & $>$~2.2   & 6   \\
 \midrule
\end{tabular}
\caption{IRAS colour index selection criteria proposed for
different objects by different authors. [x$-$y] $\equiv$
$\log(F_x)/\log(F_y)$. \textit{References}: (1) Emerson (1987), (2)
Wouterloot \& Walmsley (1986), (3) Wood \& Churchwell (1989), (4)
Braz et al. (1989), (5) P91; CMC $\equiv$ Compact Molecular Cloud,
(6) Present work (\S~\ref{sec:IRAS fluxes and color analysis}).}
 \label{tab:colour_indexes}
\end{center}
\end{table*}

\begin{figure}
  \centering
\resizebox{\hsize}{!}{  \includegraphics[angle=-90]{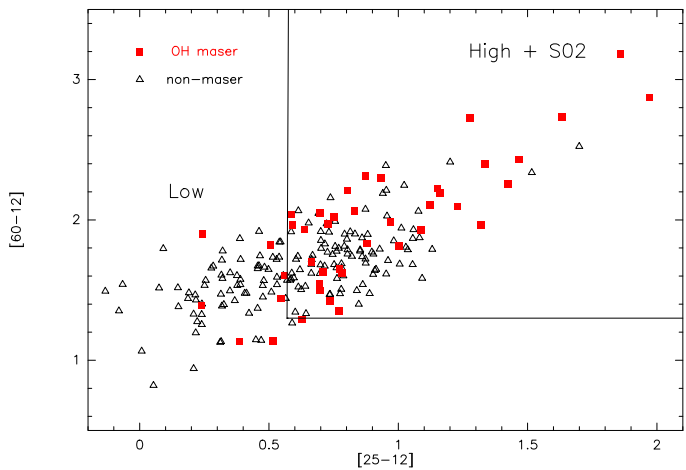}}
\caption{The [60$-$12] \textit{vs} [25$-$12] two colour plot for
the objects  searched for OH maser emission (except the seven
objects cataloged in Table 
~\ref{tab:OHmaserRes} as offset sources). The box identifies
the \textit{High} and S02 samples which contain sources which
agree with WC89 criteria to select UCHII region candidates. The
sources of the \textit{Low} sample are outside this box. OH maser
sources are plotted as filled squares, while non$-$OH masers are
shown as open triangles.}
 \label{fig:[60-25-12]}
\end{figure}

Different selection criteria have been proposed by many authors to
use the IRAS point source catalogue to identify massive young
sources (Table~\ref{tab:colour_indexes}, and references therein).
The sample of sources in this present survey is drawn from the
sample of Sridharan et al. (\citealp{sridharan02}, hereafter S02)
and Molinari et al. (\citealp{molinari96}, hereafter M96). These
two samples are believed to contain massive sources in a very
early stage of evolution prior to the forming of UCHII regions.
M96 divided their sample into two sub$-$samples, a \textit{High}
sample and a \textit{Low} sample. The S02 sample and the
\textit{High} sub$-$sample of M96 satisfy the colour selection
criteria of Wood \& Churchwell \citealp{wood89} (hereafter WC89)
for UCHII regions.  However the sources in these two samples (and
also the M96 \textit{Low} sub$-$sample) are not known to be
associated with detectable HII regions. M96 suggest that their
\textit{Low} sub$-$sample comprises objects which are in a
different evolutionary stage from those in their \textit{High}
sub$-$sample, and therefore also from the S02 sample. The two
samples (S02 and M96) have sources with flux densities brighter
than 90 Jy at 60$\mu$m. Figure~\ref{fig:[60-25-12]} shows the
[25$-$12]\footnote{[x$-$y]
  indicates $\log(F_x)/\log(F_y)$} versus [60$-$12] colour-colour
diagram, indicating the location of the \textit{High},
\textit{Low} and S02 samples, for the whole sample observed here.
Figure~\ref{fig:[60-25-12]} also shows the results of this survey
(\S~\ref{sec:results}); the detected sources of OH maser emission
are marked with filled squares.

Observations show evidence of dense molecular gas associated with
the majority of sources in the sample. The sources in the S02
sample have all been detected in CS J=2$-$1 (S02) while ammonia
was detected towards 80\% and 45\% of the \textit{High} and
\textit{Low} sub$-$samples respectively of M96. Further details
about the sources and their selection criteria can be found in S02
and M96 and references therein.

Combining the M96 and S02 samples, and accounting for the 15 sources
in common, results in a sample of 217 of HMPO candidates which have
been observed here.{ The vast majority of the sources in the sample
  have luminosities in the range $\sim10^{3}$~L$_\odot$ to
  $10^{5.5}$~L$_\odot$. However distance uncertainties affect the
  luminosity estimates of individual sources and are likely to explain
  a few sources with apparently with much lower luminosities.}
Although these objects have been previously systematically surveyed
for water and methanol masers, only a handful have previously been
searched for OH masers.

\subsection{Water and Methanol Masers in the sample}

S02 searched their sample for 22 GHz \htwoo\ and 6.7 GHz
\methanol\ masers using the Effelsberg 100~m telescope. The
detected sources (29 \htwoo\ and 26 \methanol\ masers) were mapped
with the Very Large Array (VLA) by Beuther et al.
\cite{beuther02}. The M96 sample had already been surveyed for
\htwoo\ maser by P91 with the Medicina 32~m telescope and most of
the sources were searched for 6.7~GHz \methanol\ maser emission by
Szymczak, Hrynek \& Kus (\citealp{SHK2000}) using the Toru\'{n}
32~m radio telescope. Towards the M96 sample 40 \htwoo\ and 26
\methanol\ maser sources have been reported. Therefore water and
methanol masers were found towards 36\% and 21\%, respectively, of
the 217 sources.

\subsection{Outflows in the sample}

Beuther et al. \cite{beuther02c} searched 26 sources from the S02
sample at a spatial resolution of 11$^{''}$ for bipolar molecular
outflows. The signature of outflow was found towards 21 sources of
them. The other five sources showed confusing morphology but have
strong line wings. More recently, Zhang et al. \cite{zhang05} studied
this phenomenon towards 69 sources of the M96 sample. Mapping in the CO $J
= 2$-$1$ line, they identified 39 molecular outflows.  Therefore,
among 95 sources systematically searched out of the total of 217
objects, 60 sources show clear evidence of bipolar molecular outflows.
This gives a 63~\% detection rate which indicates that molecular
outflows are very common in these regions.

\section{Observations}

\subsection{\nancay\ Observations}

Observations using the Nan\c{c}ay radio telescope were performed
between July 2002 and June 2003. At 18cm the telescope has a
beamsize of 3.5$^{'}$x19$^{'}$ (RA $\times$ Dec). We
simultaneously observed the four OH transitions at 1665, 1667,
1612 and 1720 MHz in both left and right circular polarizations.
The 8192 channel autocorrelator was configured into eight banks of
1024 channels. Each bank had a total bandwidth of 1.5625 MHz
yielding velocity resolutions of 0.284, 0.275, 0.274 and 0.266
km~\s\ at 1612, 1665, 1667 and 1720 MHz, respectively. Inband
frequency$-$switching was used during these observations. The
total integration time per sources was 18 minutes, giving a
typical noise level in a single polarisation of about 50~mJy. A
total useable velocity range of about $\pm$160 km~\s\ was covered
towards each source. The radial velocities were measured with
respect to the local standard of rest (LSR). The spectral
bandwidth was centred at the molecular gas velocity of the
observed source, as given by M96 and S02. The NAn\c{c}ay
Preprocessing Software (NAPS) program was used for the initial
data processing and eliminating bad integrations, and integrating
the whole cycles of each scan. The data were then imported into
\textit{CLASS} for further processing. In \textit{CLASS}, the
spectra were FOLDed to remove ripples coming from the
frequency$-$switching technique and finally the spectral plots
were produced.

\subsection{GBT Observations}

The Green Bank Telescope (GBT) was used to re$-$observe the sources
detected by \nancay\ in order to: (1) observe them with higher
spectral resolution; (2) decrease the contamination due to \nancay's
large beam size ($3.5^{'} \times 19^{'}$, comparing to $\sim 8^{'}$ of
GBT); and (3) make small maps to determine whether the masers were
associated with these IRAS sources or offset from them. The GBT was
also used to observe a small set of sources we did not have enough
time to observe with \nancay. The observations were carried out from
18 to 23 May 2003. The 12.5 MHz, 9 level, 8 sampler correlator setting
was used to observe all four OH lines (1665, 1667, 1612 and 1720 MHz)
in both senses of circular polarizations. After confirming the
presence of an OH maser towards the IRAS position of a source, most of
the sources were mapped with small 3 arcminute sampled maps, typically
$3\times3$ pixels in size, to determine the position of the peak
emission. For the majority of sources, where the position of the OH
masers was consistent with the IRAS position, a higher resolution
spectrum (with a velocity channel width of $\sim 0.07$ km \s) was then
obtained towards the IRAS position. The typical noise level in these
high resolution spectra as about 0.1 Jy. For the high resolution
spectra, the 12.5MHz, 3 level, 8 sampler mode was used to obtain four
times higher resolution and covering all four OH lines in both senses
of circular polarizations. Frequency switching was used during these
observation. 
The typical system temperature was $\sim$20K resulting in a typical
noise level of 50~mJy and 150~mJy in the low and high resolution
spectrum respectively.  Note that sources observed at \nancay\ but not
detected were not reobserved at GBT.

\begin{figure*}
\begin{center}
  \includegraphics[angle=-90,width=8cm]{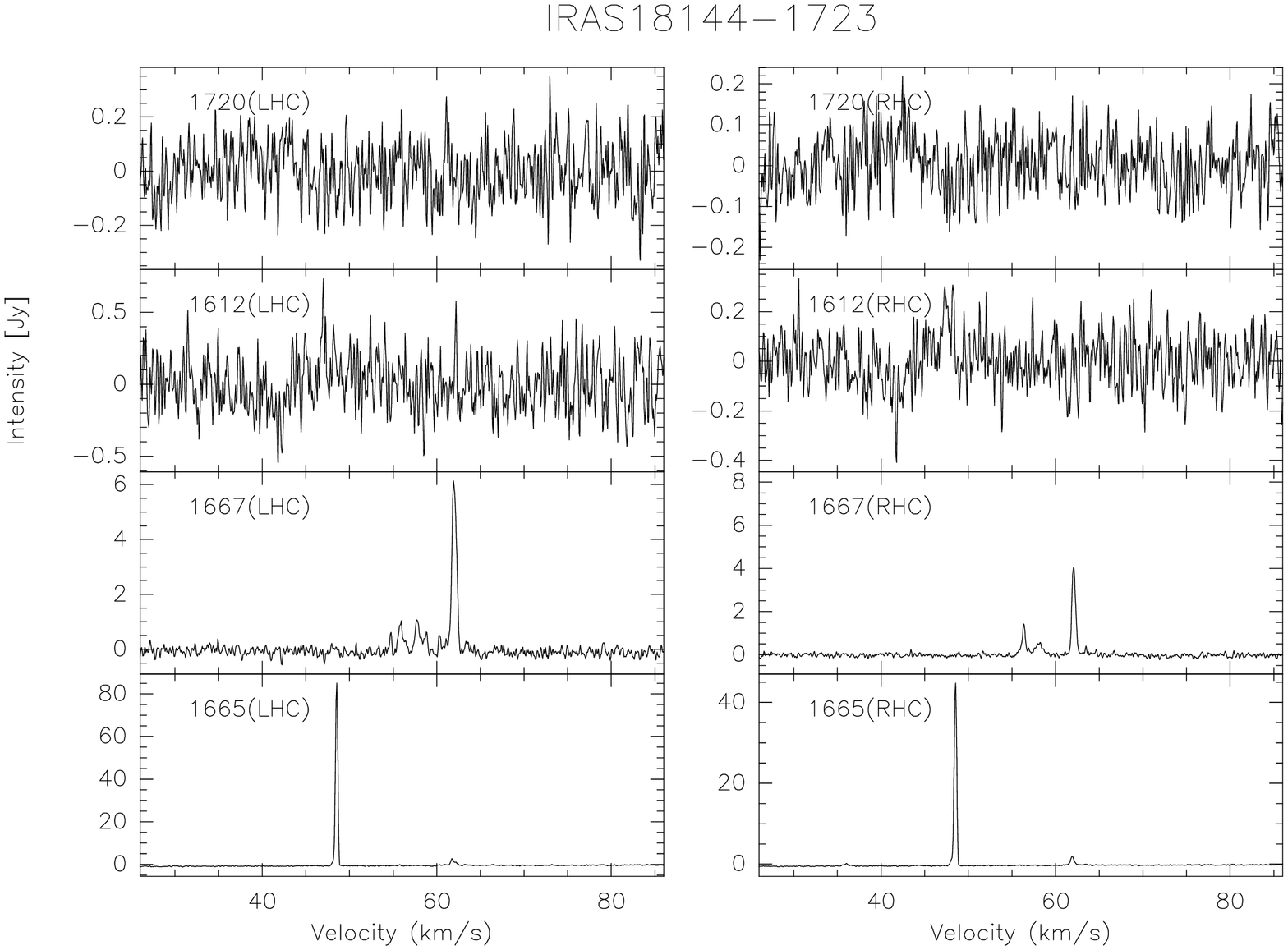}
  \includegraphics[angle=-90,width=8cm]{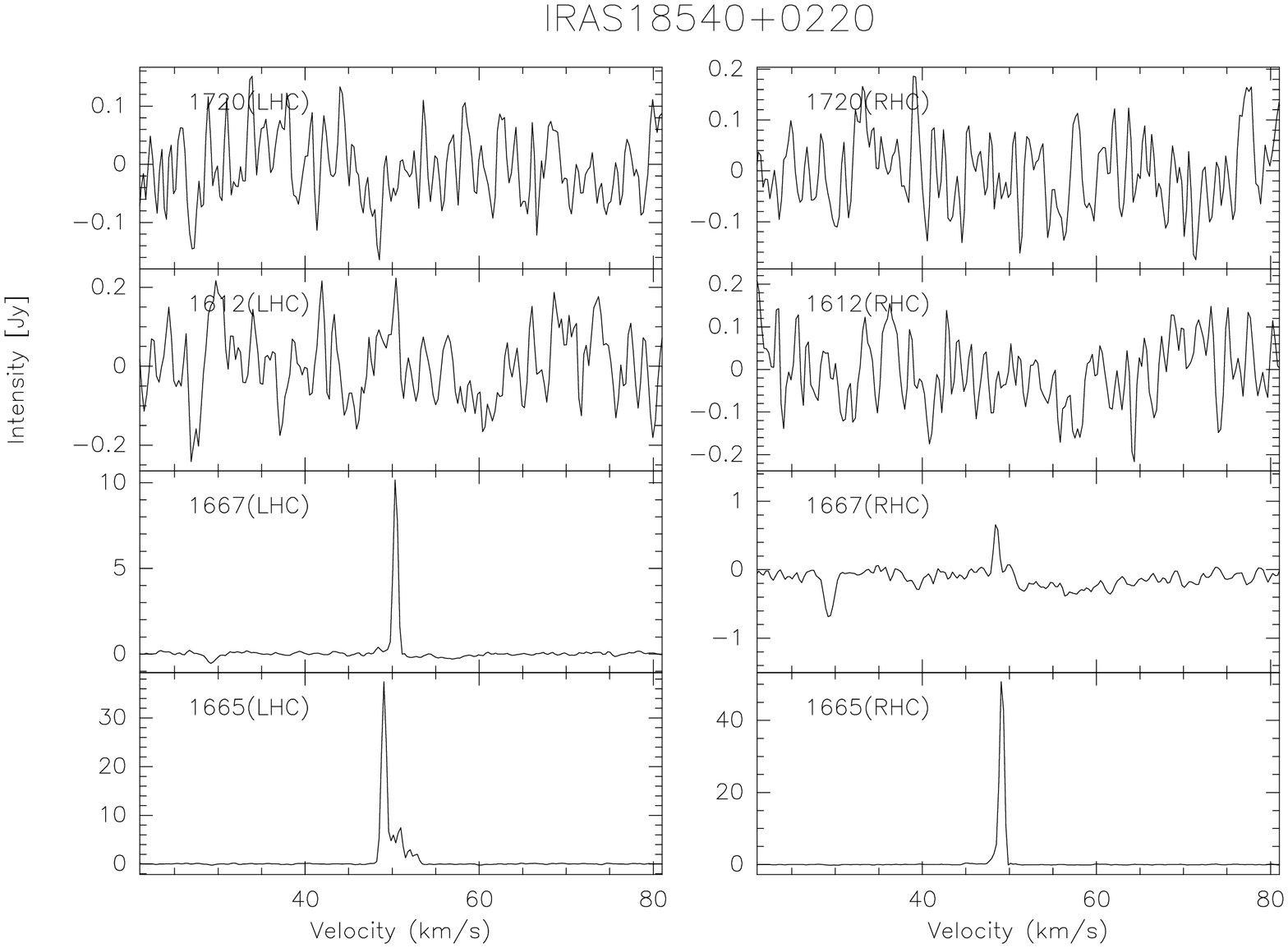}
  \includegraphics[angle=-90,width=4cm]{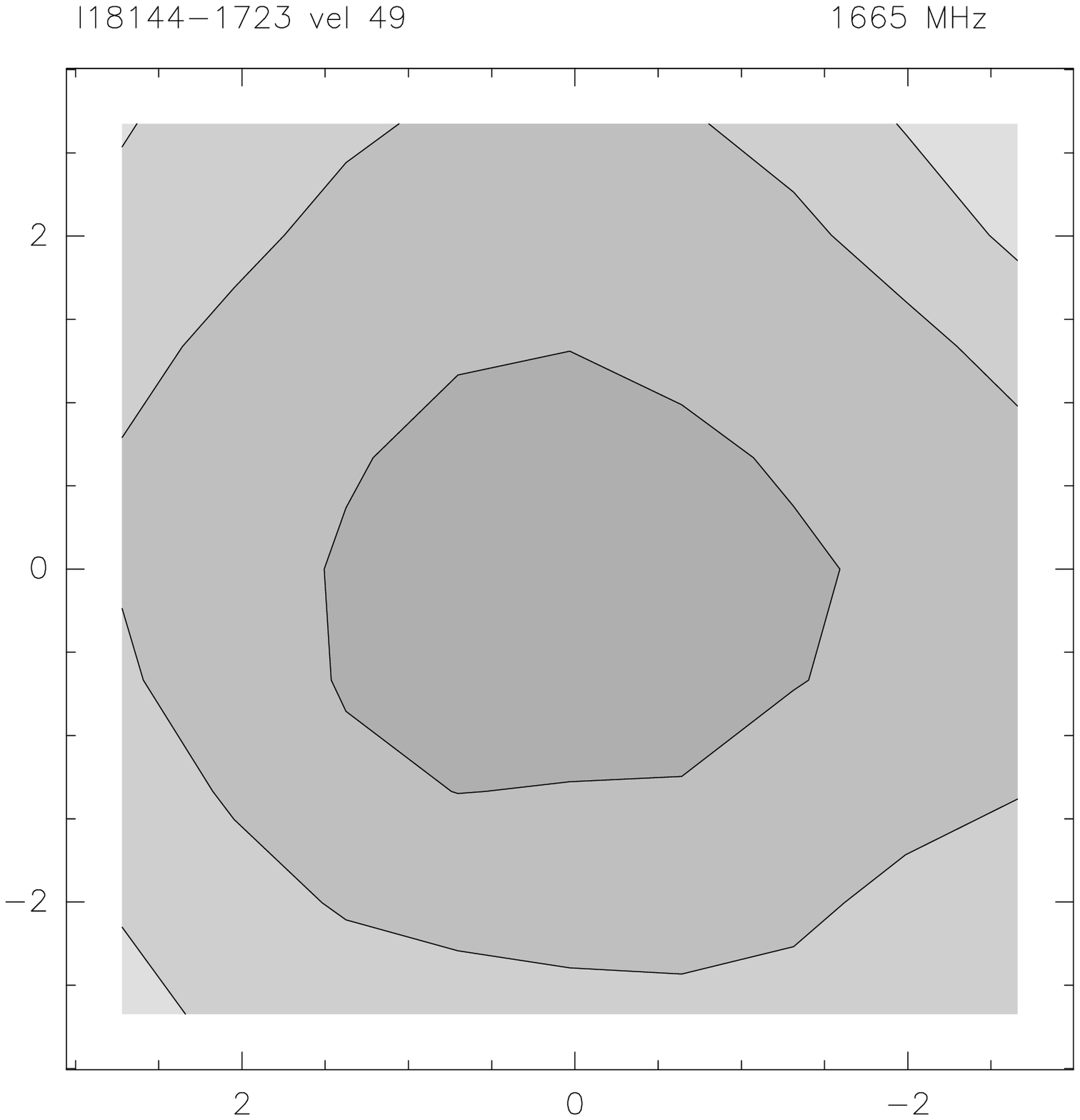}
  \includegraphics[angle=-90,width=4cm]{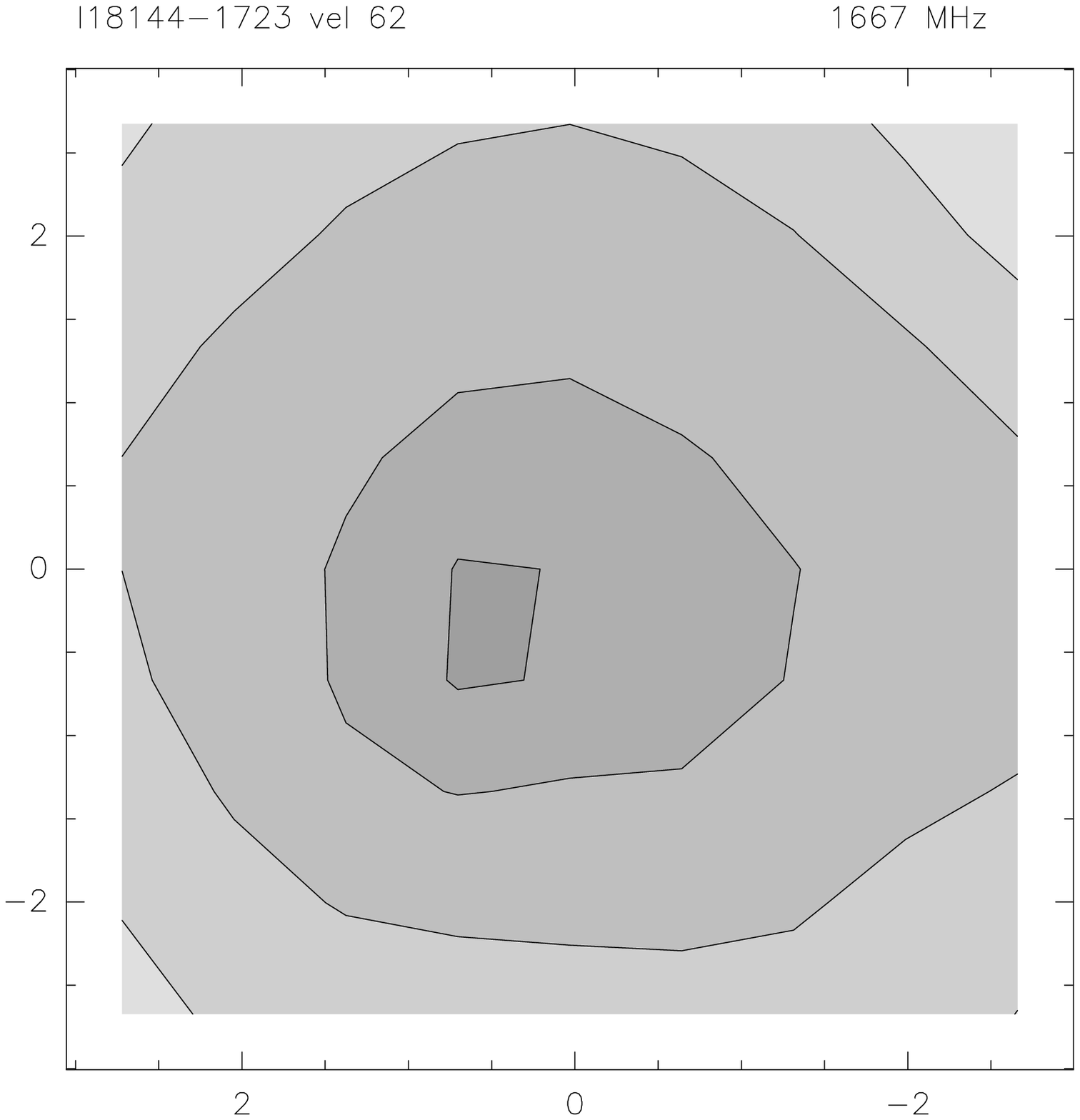}
\includegraphics[angle=-90,width=4cm]{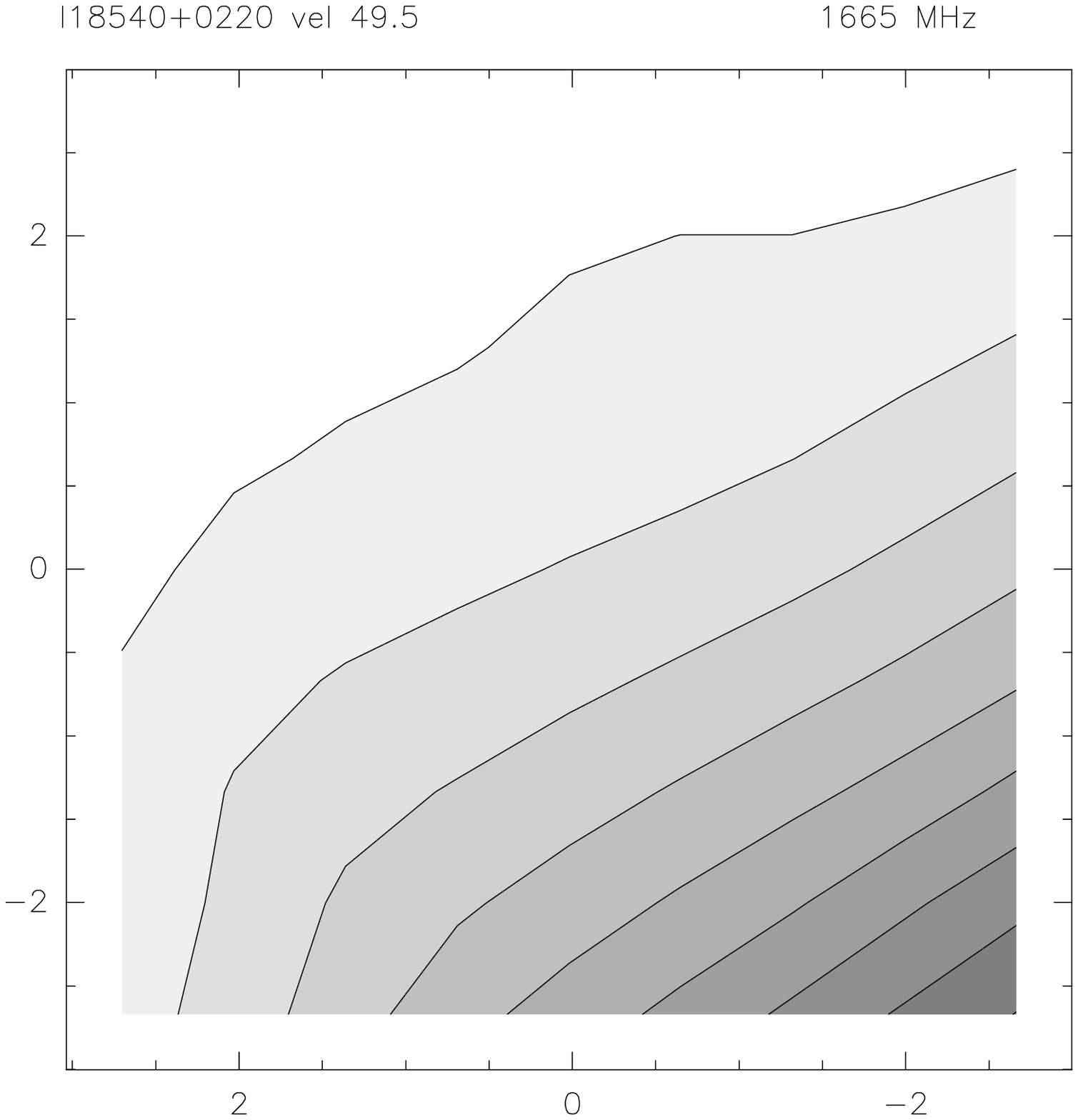}
\includegraphics[angle=-90,width=4cm]{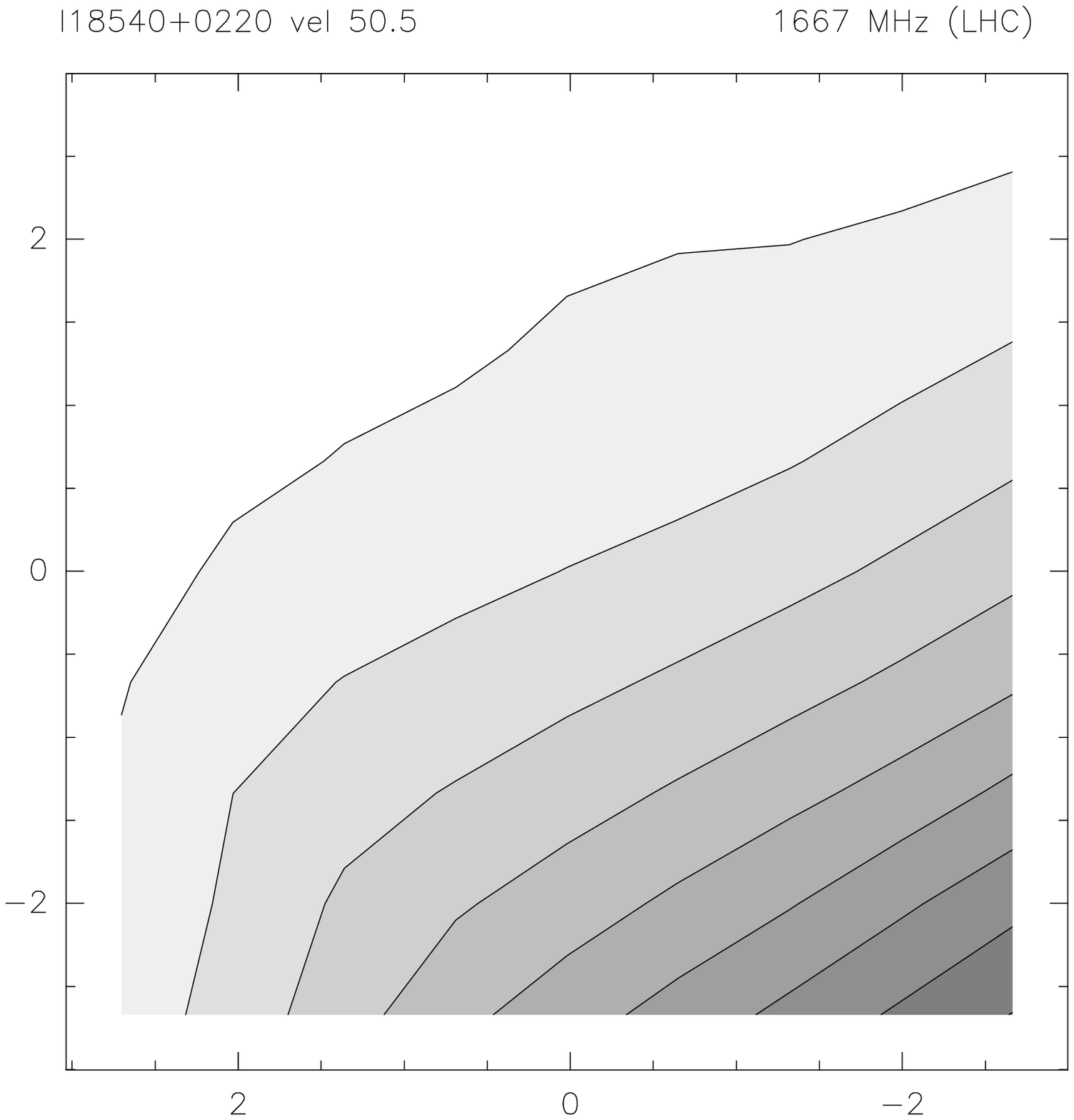}
\caption{Examples spectra and maps of OH masers detected with the GBT.
  The maps shows if the OH maser emission is consistent with the IRAS
  position (as for IRAS18144$-$1723, left panels) or offset from it
  (as for IRAS18540+0220, right panels). The maps are 9-points maps of
  the integrated intensities over the velocity range of each maser
  component. The contours range from 10\% to 90 \% of the peak flux
  given in Table~\ref{tab:OHmaserRes} in steps of 10\%. The axes show
  RA and Dec offset from the IRAS position in arcminutes.}
 \label{fig:GBT-spec-maps-s}
\end{center}
\end{figure*}

\section{Results}
\label{sec:results}

Combining the results from both \nancay\ and GBT, a total of 63
sources show OH maser emission, defined as bright and narrow
polarized lines, in one or more of the OH transitions. Of these 36
have not been reported before.

The nature of the OH maser emission provides the opportunity to
distinguish between sources which have same colours as HMPOs but
are in fact late type stars around which OH masers are also known
to occur (e.g. Cohen 1989). Towards evolved stars the 1612 MHz
line spectrum usually shows double-peaked profile with sharp
external edges and smooth internal ones with the two peaks 
separated by 15 to 40 km s$^{-1}$ (Cohen 1989).  Of the sources
detected here 7 have 1612 MHz masers which indicate that they are
OH/IR candidates.

The 63 sources detected include 57 detected at both \nancay\ and GBT,
and 6 which were detected at \nancay\ but not observed at the GBT.  We
categorize these 6 sources as \textit{not confirmed}, and do not
consider them in our subsequent analysis.

The association of the maser emission and the IRAS source that we
searched towards is confirmed by GBT maps for the first 46 sources in
the table. Only these 46 objects are considered in the IRAS-related
statistical analysis that follows. Thirty nine of these
IRAS-associated sources are typical star-forming regions with masers
strong in the main lines, 1665 and 1667 MHz. One source, IRAS
18463+0052, is associated with an OH/IR star. The remaining 6 sources
have OH masers which are not typical of star-forming regions; they
show maser emission in one of the satellite lines only (see
\S~\ref{sec:1720}). There are remaining 11 confirmed sources with OH
maser emission offset by $>$ 2$^{'}$ from the IRAS position are
cataloged as {\it offset} sources in Table~\ref{tab:OHmaserRes}. They
include 4 OH/IR candidates and 7 sources with OH spectra
characteristic of star-forming regions.

The 57 confirmed maser sources and their OH line parameters are
listed in Table~\ref{tab:OHmaserRes}. Column 1 gives the IRAS name
of the source, columns 2 and 3 the maser position, where measured,
with the uncertainties, column 4 the frequency and circular
polarization of the maser line, with either L or R referring to
lefthand or righthand circular polarisation respectively, column 5
the feature central velocity (relative to the  LSR), columns 6 and
7 the velocity interval at zero intensity, column 8 the peak flux
density in Jy, column 9 rms noise, column 10 the ratio of the
\methanol\ peak flux density to the OH peak flux density, R, as
defined by Caswell (1998), and column 11 some comments.

Figure~\ref{fig:GBT-spec-maps-s} shows an example set of OH maser
spectra and map from the GBT (IRAS18144--1723, left panels). The
figure also shows an example of one of the sources cataloged as {\it
  offset} because their map reveal that the emission is not consistent
with the IRAS position (IRAS18540+0220, right panels).  Many of the
sources with maser emission, for example IRAS 17527--2339
(Figure~\ref{fig:spectra}), show `conjugate' behaviour with one of the
satellite lines in emission while the other is in absorption (Elitzur
\citealp{Elitzur76}).  The maps and spectra of all the OH detected
sources are shown in Figure~\ref{fig:GBT-map} and
Figure~\ref{fig:spectra} respectively.  The maps and spectra of the
sources with maser emission offset from the IRAS position by $>$
2$^{'}$ are separated in to Figure~\ref{fig:GBT-map-offset} and
Figure~\ref{fig:spectra-offset} respectively.

A further 79 sources were detected in thermal emission and/or
absorption.  They are listed in Table~\ref{S_no_detection} along with
the 75 sources which were not detected in our observations.
Table~\ref{S_no_detection} also list the 6 sources categorized as
\textit{not confirmed} maser sources.  A brief description and
discussion of each of the OH maser sources is given in
Appendix~\ref{app:sources}.

\onecolumn
{\scriptsize%
  \begin{center}
    \tablefirsthead{
      \toprule%
      \toprule%
      IRAS Name                              &
      \multicolumn{2}{c}{position}           &
      \multicolumn{1}{c}{Frequency}          &
      \multicolumn{3}{c}{Velocity at}           &
      \multicolumn{1}{c}{Flux at}               &
      \multicolumn{1}{c}{RMS}                &
      R                                      &
      Ref./Notes                             \\
      \cmidrule(r){2-3}\cmidrule(lr){5-7}
      &
      \multicolumn{1}{c}{RA(J2000)}          &
      \multicolumn{1}{c}{DEC(J2000)}         &
      &
      \multicolumn{1}{c}{$V_{peak}$}      &
      \multicolumn{1}{c}{$V_{min}$}     &
      \multicolumn{1}{c}{$V_{max}$}       &
      \multicolumn{1}{c}{$S_{peak}$}      &
      &
      &
      \\
      \cmidrule(r){5-7}
      &
      \multicolumn{1}{c}{h~~m~~s}                           &
      \multicolumn{1}{c}{$^{\circ}$~~$'$~~$''$}         &
      &
      \multicolumn{3}{c}{km \s}        &
      \multicolumn{1}{c}{Jy}         &
      \multicolumn{1}{c}{Jy}         &
      &
      \\
      \midrule
    }
    \tablehead{%
      \toprule%
      \multicolumn{10}{l}{\small\sl continued from previous page}\\
      \toprule%
      IRAS Name                                 &
      \multicolumn{2}{c}{position}       &
      \multicolumn{1}{c}{Frequency}         &
      \multicolumn{3}{c}{Velocity at}         &
      \multicolumn{1}{c}{Flux at}         &
      \multicolumn{1}{c}{RMS}         &
      R                                      &
      Ref./Notes                            \\
      \cmidrule(r){2-3}\cmidrule(lr){5-7}
      &
      \multicolumn{1}{c}{RA(J2000)}       &
      \multicolumn{1}{c}{DEC(J2000)}         &
      &
      \multicolumn{1}{c}{$V_{peak}$}         &
      \multicolumn{1}{c}{$V_{min}$}         &
      \multicolumn{1}{c}{$V_{max}$}         &
      \multicolumn{1}{c}{$S_{peak}$}         &
      &
      &
      \\
      \cmidrule(r){5-7}
      &
      \multicolumn{1}{c}{h~~m~~s}                           &
      \multicolumn{1}{c}{$^{\circ}$ ~ $'$ ~ $''$}         &
      &
      \multicolumn{3}{c}{km \s}        &
      \multicolumn{1}{c}{Jy}         &
      \multicolumn{1}{c}{Jy}         &
      &
      \\
      \midrule
    }
    \tabletail{%
      \midrule
      \multicolumn{10}{r}{\small\sl continued on next page}\\
      \midrule}
    \tablelasttail{\bottomrule}
  \tablecaption{The detected OH maser lines and their
parameters. The \textit{offset} star forming region and OH/IR
candidates are separated at the end of this table. The listed
positions are measured from GBT observations except some position
taken from higher resolution observations of Argon et al.
\cite{argon00} and Edris et al. \cite{Edris05} (Ref. 1 and 2 in
the table respectively).}
\begin{supertabular}{lcccccccccc}

\object{05137+3919} &    05 17 12.8 $\pm$ 2.2    &       39 22 05  $\pm$  38      &   1665R    & $-$21.58   &  $-$24.5   & $-$21.2    &  2.20  & 0.40 &  &  \\
\object{05274+3345} &    05 31 06.4 $\pm$ 1.4    &       33 47 27  $\pm$  19      &   1665R    &  $-$3.61   &   $-$5     &   0.6    &  1.34  & 0.18 & 70 & \\
           &                             &                                   &   1667L    &  $-$4.93   &          &          &  0.14  & 0.04 &  &  \\
           &                             &                                   &   1612L    &  $-$3.87   &          &          &  0.13  & 0.05 &  &  \\
\object{05358+3543} &    05 39 13.0 $\pm$ 0.1    &       35 45 51   $\pm$  01      &   1665L    & $-$10.88   &  $-$16.5   &  $-$8.5    &  2.82  & 0.47 & 91 & 1   \\
           &                             &                                   &   1667L    & $-$10.53   &          &          &  0.93  & 0.17 &   & \\
\object{05382+3547} &    05 41 12.9 $\pm$ 1.3    &       35 54 06  $\pm$  20      &   1665R    & $-$26.83   &  $-$27.0   & $-$26.5    &  0.50  & 0.13 & 15   \\
\object{06056+2131} &    06 08 52.4 $\pm$ 1.6    &       21 34 06  $\pm$  23      &   1665L    &  10.14   &    3.0   &  11.0    &  3.23  & 0.51 & 6 &  \\
           &                             &                                   &   1667R    &   9.44   &          &          &  0.22  & 0.05 &  & off source \\
           &                             &                                   &   1720R    &   3.37   &          &          &  0.60  & 0.10 &  &   \\
\object{17527$-$2439} &    17 55 28.3 $\pm$ 1.9    &      $-$24 36 36  $\pm$  27      &   1665R    &  11.53   &    8.2   &  12.2    &  0.40  & 0.14 & &    \\
\object{18018$-$2426} &    18 04 53.1 $\pm$ 0.1    &      $-$24 26 41   $\pm$  01      &   1665R    &  10.84   &   10.0   &  12.0    &  8.10  & 1.83 &  &  \\
           &                             &                                   &   1667L    &  11.05   &          &          &  0.98  & 0.20 &  & 1   \\
\object{18024$-$2119} &    18 05 25.6 $\pm$ 2.3    &      $-$21 14 59  $\pm$  19      &   1665R    &  $-$4.05   &   $-$9.0   &  31.0    &  0.69  & 0.17 & 145 & \\
           &                             &                                   &   1667R    &  $-$3.75   &          &          &  0.38  & 0.10 &  &  \\
\object{18048$-$2019} &    18 07 44.6 $\pm$ 2.2    &      $-$20 18 41  $\pm$  38      &   1665R    &  44.36   &   40.0   &  44.7    &  0.34  & 0.08 & 104 &  \\
           &                             &                                   &   1667L    &  40.28   &          &          &  0.27  & 0.09 &  &  \\
\object{18089$-$1732} &    18 11 51.4 $\pm$ 0.1    &      $-$17 31 29   $\pm$  01      &   1665L    &  32.92   &   31.0   &  36.0    &  10.30 & 1.86 & 6 & 1   \\
           &                             &                                   &   1667L    &  33.36   &          &          &   2.00 & 0.32 &  &  \\
\object{18090$-$1832} &    18 11 47.4 $\pm$ 1.5    &      $-$18 29 47  $\pm$  26      &   1665R    & 108.9    &  103.0   & 110.0    &   0.70 & 0.08 & 110 &    \\
           &                             &                                   &   1667R    & 106.6    &          &          &   0.49 & 0.08 &  &  \\
\object{18102$-$1800} &    18 13 04.4 $\pm$ 1.5    &      $-$18 00 23  $\pm$  16      &   1665R    &  24.40   &   24.0   &  25.0    &   0.42 & 0.10 & 31 &    \\
\object{18144$-$1723} &    18 17 26.5 $\pm$ 1.1    &      $-$17 22 29  $\pm$  16      &   1665L    &  48.33   &   48.0   &  64.0    &  35.9  & 3.45 & 1 & \\
             &                             &                                   &   1667L    &  61.75   &          &          &   4.80 & 0.62 &  &  \\
\object{18182$-$1433} &    18 21 11.0 $\pm$ 1.0    &      $-$14 31 23   $\pm$  14      &   1665L    &  61.55   &   58.0   &  64.0    &   0.72 & 0.16 & 33 &    \\
           &                             &                                   &   1667L    &  62.45   &          &          &   0.40 & 0.12 &  &  \\
\object{18236$-$1205} &    18 26 36.2 $\pm$ 1.0    &      $-$12 04 54   $\pm$  14      &   1665R    &  31.09   &   20.0   &  32.0    &   0.80 & 0.16 & 8 &    \\
           &    18 26 31.7 $\pm$ 1.0    &      $-$12 03 26   $\pm$  14      &   1667L    &  62.45   &          &          &   0.28 & 0.06 &  &  \\
\object{18264$-$1152} &    18 29 19.2 $\pm$ 1.1    &      $-$11 50 05  $\pm$  17      &   1720L    &  39.31   &   37.5   &  43.5    &   2.30 & 0.55 &  &  \\
\object{18278$-$1009} &    18 30 37.9 $\pm$ 1.1    &      $-$10 07 25  $\pm$  17      &   1665R    & 119.7    &  118.0   & 121.0    &   0.42 & 0.09 & 33 &  \\
\object{18290$-$0924} &    18 31 46.4 $\pm$ 1.1    &      $-$09 22 14  $\pm$  15      &   1665R    &  78.33   &   76.0   &  84.0    &   2.30 & 0.41 & 5 &  \\
           &                             &                                       &   1667R    &  78.45   &          &          &   0.34 & 0.90 &  &  \\
\object{18310$-$0825} &    18 33 36.0 $\pm$ 0.8    &      $-$08 19 46  $\pm$  13      &   1667L    &  88.74   &   88.0   &  89.0    &   1.40 & 0.21 &  &    \\
\object{18316$-$0602} &    18 34 25.9 $\pm$ 1.1    &      $-$06 00 01  $\pm$  16      &   1665R    &  39.90   &   36.0   &  46.0    &   6.00 & 0.87 & 30 &  \\
           &                             &                                   &   1667L    &  40.41   &          &          &   3.38 & 0.48 &  &  \\
\object{18345$-$0641} &    18 47 08.0 $\pm$ 2.8    &      $-$02 19 05  $\pm$  40      &   1612R    &  93.52   &   92.0   &  96.0    &   0.43 & 0.07 & 23 &    \\
\object{18360$-$0537} &    18 38 42.2 $\pm$ 1.1    &      $-$05 36 25  $\pm$  16      &   1665R    & 102.9    &  102.0   & 106.0    &   0.54 & 0.05 &  &  \\
           &    18 38 47.2 $\pm$ 1.2    &      $-$05 35 40  $\pm$  17      &   1667L    & 105.3    &          &          &   0.74 & 0.05 &  &  \\
\object{18385$-$0512} &    18 41 18.2 $\pm$ 1.0    &      $-$05 08 57   $\pm$  14      &   1665R    &  24.87   &   21.0   &  30.0    &   1.01 & 0.04 & &   \\
\object{18440$-$0148} &    18 46 37.8 $\pm$ 1.0    &      $-$01 44 27   $\pm$  13      &   1665R    & 101.4    &   99.0   & 110.0    &   6.00 & 0.08 & 0.5 &   \\
             &                             &                                     &   1667R    & 102.6    &          &          &   2.00 & 0.10 &  &  \\
\object{18454$-$0158} &  18 48 01.3 $\pm$ 1.0    &      $-$01 54 34  $\pm$  15    &   1665L    &  39.6    &   30.0   &  44.0    &   0.35 & 0.10 &  &  \\
\object{18463+0052} &    18 48 46.8 $\pm$ 2.2     &       00 56 55  $\pm$  24      &   1612R    &  67.32   &   67.0   &  70.0    &   2.26 & 0.09 &  & OH/IR \\
\object{18488+0000} &    18 51 30.5 $\pm$ 1.0    &       00 03 21   $\pm$  16      &   1665R    &  79.57   &   79.0   &  87.0    &   4.52 & 0.07 & 6 &  \\
           &                             &                                   &   1667R    &  77.90   &          &          &   2.66 & 0.05 &  &  \\
\object{18507+0121} &    18 53 18.2 $\pm$ 1.2    &       01 24 30   $\pm$  18      &   1665L    &  55.78   &   53.0   &  56.0    &   2.00 & 0.08 & 14 &  \\
           &                             &                                   &   1667L    &  53.88   &          &          &   1.10 & 0.09 &  &  \\
\object{18527+0301} &    18 54 46.5 $\pm$ 2.5    &       03 05 07   $\pm$  40      &   1665R    &  74.44   &   72.0   &  75.0    &   0.21 & 0.02 & 48 &  \\
           &                             &                                   &   1667L    &  73.30   &          &          &   0.16 & 0.02 &  &  \\
\object{18553+0414} &    18 57 50.7 $\pm$ 1.2    &       04 18 36  $\pm$  18      &   1720L    &   6.63   &    4.0   &   8.0    &   0.67 & 0.03 &  &  \\
\object{18566+0408} &    18 59 08.7 $\pm$ 1.1    &       04 10 21  $\pm$  17      &   1665L    &  83.41   &   52.0   &  92.0    &   1.00 & 0.12 & 7 &  \\
           &    18 59 10.4 $\pm$ 1.0    &       04 13 21   $\pm$  14      &   1667L    &  81.52   &          &          &   0.50 & 0.14 &  &  \\
\object{19035+0641} &    19 06 01.6 $\pm$ 0.0    &       06 46 35   $\pm$  01      &   1665L    &  32.44   &   24.2   &  36.2    &  90.00 & 0.61 & 0.2 & 1  \\
           &                             &                                   &   1667R    &  27.33   &          &          &  22.30 & 0.27 &  &  \\
\object{19092+0841} &    19 11 45.9 $\pm$ 0.4    &       08 46 49  $\pm$  06      &   1665R    &  57.87   &   54.0   &  62.0    &   3.45 & 0.04 & 3 &  \\
           &    19 11 46.6 $\pm$ 0.4    &       08 46 19  $\pm$  10      &   1667L    &  60.51   &          &          &   1.75 & 0.04 &  &  \\
\object{19118+0945} &    19 14 29.7 $\pm$ 1.6    &       09 51 47  $\pm$  46      &   1665R    &  61.25   &   61.0   &  71.0    &   0.35 & 0.05 &  &  \\
           &                             &                                   &   1667L    &  58.90   &          &          &   0.22 & 0.05 &  &  \\
\object{19217+1651} &    19 23 57.9 $\pm$ 0.9    &       16 56 42  $\pm$  13      &   1665L    &   0.21   &   $-$2.0 &  10.0    &   1.35 & 0.05 & 1 &  \\
           &                             &                                   &   1667L    &   6.69   &          &          &   1.22 & 0.05 &  &  \\
\object{19220+1432} &    19 24 19.7 $\pm$ 1.5    &       14 37 23  $\pm$  30      &   1720R    &  60.67   &   59.5   &  61.0    &   0.44 & 0.08 &  &  \\
\object{19374+2352} &    19 39 37.4 $\pm$ 4.0    &       23 59 53   $\pm$ 109      &   1665R    &  37.06   &   35.0   &  40.0    &   0.46 & 0.06 &  &  \\
           &                             &                                   &   1667R    &  37.00   &          &          &   0.13 & 0.02 &  &  \\
           &                             &                                   &   1720R    &  37.05   &          &          &   0.17 & 0.03 &  &  \\
\object{19388+2357} &   19 41 10.2  $\pm$ 2.2    &       24 03 44  $\pm$  25      &   1665L    &  35.81   &   34.0   &  39.0    &   0.32 & 0.04 & 77 &  \\
\object{19410+2336} &   19 43 12.2  $\pm$ 1.1    &       23 44 03  $\pm$  12      &   1665L    &  20.67   &   20.0   &  23.0    &   0.71 & 0.03 & 48 &  \\
\object{20062+3550} &   20 08 12.7  $\pm$ 1.2    &       35 59 20  $\pm$  20      &   1665R    &   0.77   &   $-$1.0   &   2.2    &   0.17 & 0.03 & 59 &    \\
\object{20126+4104} &   20 14 26.06  $\pm$ 0.002    &       41 13 32.63  $\pm$  0.02      &   1665R    & $-$12.27   &  $-$16.0   &   2.0    &   2.37 & 0.11 & 16 &  2  \\
\object{20188+3928} &   20 20 41.8  $\pm$ 1.0    &       39 37 42   $\pm$  12      &   1720L    &  $-$1.03   &   $-$3.0   &   4.0    &   4.18 & 0.14 & &   \\
\object{20227+4154} &   20 24 34.6  $\pm$ 1.3    &       42 06 12  $\pm$  22      &   1665L    &  24.29   &   10.0   &  25.0    &   0.80 & 0.07 &  &  \\
\object{22198+6336} &   22 21 00.7  $\pm$ 1.3    &       63 51 57  $\pm$  16      &   1665L    & $-$12.14   &  $-$23.0   & $-$10.0    &   2.00 & 0.08 &  &    \\
           &   22 21 37.6  $\pm$ 0.4    &       63 51 47  $\pm$  08      &   1667L    & $-$13.00   &          &          &   3.50 & 0.17 &  &  \\
\object{22272+6358} &   22 28 58.5  $\pm$ 0.8    &       64 15 52  $\pm$  14      &   1665L    & $-$12.12   &  $-$12.0   &  $-$8.0    &   1.10 & 0.06 & 83 &    \\
           &                             &                                   &   1667L    & $-$11.96   &          &          &   1.68 & 0.07 &  &  \\
\object{23139+5939} &   23 16 06.8  $\pm$ 1.0    &       59 58 52   $\pm$  13      &   1612R    & $-$72.93   &  $-$74.0   & $-$68.0    &   0.63 & 0.11 &  &    \\
      \midrule
      \multicolumn{9}{l}{\small\sl Masers sources offset}\\
      \cmidrule(l){1-2}
\object{04579+4703}  &                             &                                   &   1720R    & $-$17.70 &  $-$18.1 & $-$16.8  &  0.22  & 0.05 &    \\
\object{06382+0939}  &                             &                                   &   1665R    &  11.48   &    9.2   &  13.0    &  0.80  & 0.16 &  &  \\
\object{18408$-$0348}&                             &                                   &   1665R    &  89.60   &   88.0   & 113.0    &  1.00  & 0.10 &  & \\
\object{18511+0146}  &                             &                                   &   1665R    &  47.91   &   47.0   &  49.0    &  0.38  & 0.04 &  & \\
\object{18540+0220}  &                             &                                   &   1665R    &  49.13   &   48.0   &  53.5    & 53.73  & 0.19 &  & \\
                     &                             &                                   &   1667L    &  50.38   &          &          & 10.57  & 0.09 &  & \\
\object{18586+0106}  &                             &                                   &   1665L    &  42.00   &   38.5   &  46.0    &   0.50 & 0.18 &  & \\
                     &                             &                                   &   1667L    &  42.50   &          &          &   0.70 & 0.17 &  & \\
                     &                             &                                   &   1720R    &  38.81   &          &          &   3.00 & 0.31 &  &  \\
\object{20099+3640}  &                             &                                   &   1665R    & $-$42.74 &$-$43.0   & $-$42.0  &  0.30  & 0.04 &  & \\
      \midrule
      \multicolumn{9}{l}{\small\sl OH/IR sources offset}\\
      \cmidrule(l){1-2}
\object{18258$-$0737} &                             &                                   &   1612R    &  93.00   &   63.0   &  95.0    &  2.00  & 0.10 &  & OH/IR \\
\object{18348$-$0616} &                             &                                   &   1612L    &  70.00   &   25.0   &  73.0    &  1.50  & 0.18 &  & OH/IR \\
\object{18424$-$0329} &                             &                                   &   1612L    &  40.00   &   38.0   &  82.0    &  0.50  & 0.13 &  & OH/IR \\
\object{18565+0349}   &                             &                                   &   1612L    &  27.00   &   25.0   &  62.0    &  1.50  & 0.11 &  & OH/IR \\
   \bottomrule
    \end{supertabular}
 \label{tab:OHmaserRes}
  \end{center}
}
\twocolumn

\begin{table}
  \begin{center}
\caption{Non-OH maser sources cataloged as: sources with
thermal emission and/or absorption in one or more of the OH maser
lines, sources with no detectable features, not confirmed maser
sources (sources detected with \nancay\ only), not confirmed OH/IR
sources.} 
\label{S_no_detection}
\begin{tabular}{llll}
      \toprule%
      \toprule%
      \multicolumn{4}{l}{\small\sl IRAS sources with thermal absorption and/or thermal emission}\\
      \cmidrule(l){1-4}
  \object{00420+5530}  &  \object{18167$-$1614}&  \object{18551+0302}  &  \object{20081+2720}  \\
  \object{05168+3634}  &  \object{18172$-$1548}&  \object{18567+0700}  &  \object{20180+3558}   \\
  \object{05345+3157}  &  \object{18212$-$1320}&  \object{18571+0326}  &  \object{20205+3948}   \\
  \object{05373+2349}  &  \object{18223$-$1243}&  \object{18571+0349}  &  \object{20216+4107}   \\
  \object{06061+2151}  &  \object{18256$-$0742}&  \object{18596+0536}  &  \object{20217+3947}   \\
  \object{06063+2040}  &  \object{18288$-$0158}&  \object{19001+0402}  &  \object{20220+3728}   \\
  \object{06105+1756}  &  \object{18306$-$0835}&  \object{19002+0454}  &  \object{20293+3952}   \\
  \object{06299+1011}  &  \object{18337$-$0743}&  \object{19012+0505}  &  \object{20293+4007}   \\
  \object{06303+1021}  &  \object{18363$-$0554}&  \object{19012+0536}  &  \object{20319+3958}   \\
  \object{17417$-$2851}&  \object{18396$-$0431}&  \object{19023+0538}  &  \object{20332+4124}    \\
  \object{17450$-$2742}&  \object{18426$-$0204}&  \object{19045+0518}  &  \object{20333+4102}    \\
  \object{17495$-$2624}&  \object{18437$-$0216}&  \object{19088+0902}  &  \object{20343+4129}    \\
  \object{17582$-$2234}&  \object{18445$-$0222}&  \object{19175+1357}  &  \object{21078+5211}   \\
  \object{18024$-$2231}&  \object{18447$-$0229}&  \object{19198+1423}  &  \object{21548+5747}   \\
  \object{18039$-$2052}&  \object{18454$-$0136}&  \object{19295+1637}  &  \object{22551+6221}    \\
  \object{18134$-$1942}&  \object{18517+0437}  &  \object{19332+2028}  &  \object{23026+5948}   \\
  \object{18156$-$1343}&  \object{18530+0215}  &  \object{19343+2026}  &  \object{23033+5951}   \\
  \object{18159$-$1550}&  \object{18532+0047}  &  \object{19368+2239}  &  \object{23140+6121}   \\
  \object{18159$-$1648}&  \object{18537+0145}  &  \object{20050+2720}  &  \object{23545+6508}   \\
  \object{18162$-$1612}&  \object{18544+0112}  &  \object{20051+3435}  & \\

      \midrule
      \multicolumn{4}{l}{\small\sl IRAS sources with no detected absorption or emission}\\
      \cmidrule(l){1-3}
  \object{00117+6412}  &  \object{18272$-$1217}&  \object{19542+3004}  & \object{22187+5559}   \\
  \object{01420+6401}  &  \object{18317$-$0513}&  \object{20028+2903}  & \object{22267+6244}   \\
  \object{03211+5446}  &  \object{18355$-$0550}&  \object{20056+3350}  & \object{22305+5803}   \\
  \object{04034+5116}  &  \object{18372$-$0541}&  \object{20106+3545}  & \object{22344+5909}   \\
  \object{05490+2658}  &  \object{18470$-$0044}&  \object{20278+3521}  & \object{22457+5751}   \\
  \object{05553+1631}  &  \object{18472$-$0022}&  \object{20286+4105}  & \object{22506+5944}   \\
  \object{06068+2030}  &  \object{18521+0134}  &  \object{20321+4112}  & \object{22570+5912}   \\
  \object{06103+1523}  &  \object{19043+0726}  &  \object{20406+4555}  & \object{23146+5954}   \\
  \object{06104+1524}  &  \object{19077+0839}  &  \object{20444+4629}  & \object{23151+5912}   \\
  \object{06155+2319}  &  \object{19094+0944}  &  \object{21046+5110}  & \object{23152+6034}   \\
  \object{06291+0421}  &  \object{19183+1556}  &  \object{21080+4758}  & \object{23314+6033}   \\
  \object{06308+0402}  &  \object{19045+0813}  &  \object{21202+5157}  & \object{23330+6437}   \\
  \object{06584$-$0852}&  \object{19213+1723}  &  \object{21307+5049}  & \object{23385+6053}   \\
  \object{17571$-$2328}&  \object{19266+1745}  &  \object{21336+5333}  & \object{23448+6010}   \\
  \object{17504$-$2519}&  \object{19282+1814}  &  \object{21391+5802}  & \object{23507+6230}   \\
  \object{18014$-$2428}&  \object{19403+2258}  &  \object{21519+5613}  & \\
  \object{18123$-$1203}&  \object{19411+2306}  &  \object{21526+5728}  & \\
  \object{18151$-$1208}&  \object{19413+2332}  &  \object{22134+5834}  & \\
  \object{18197$-$1351}&  \object{19458+2442}  &  \object{22147+5948}  & \\
  \object{18247$-$1147}&  \object{19471+2641}  &  \object{22172+5549}  & \\
      \midrule
      \multicolumn{4}{l}{\small\sl Maser sources not confirmed}\\
      \cmidrule(l){1-2}
  \object{00070+6503}  &  \object{18311$-$0701} &  \object{18431$-$0312} &  \object{19074+0752}   \\
      \midrule
      \multicolumn{4}{l}{\small\sl OH/IR sources not confirmed}\\
      \cmidrule(l){1-2}
  \object{18308$-$0841} &  \object{18460$-$0307}  \\
   \bottomrule
   \bottomrule
\end{tabular}
  \end{center}
\end{table}

\section{Detection Statistics and Analysis}

The survey observations detected 63 OH maser sources out of 217
IRAS sources. For 46 of these the association with the IRAS source
is confirmed by maps made with the GBT. Of these 36 are new
detections. That gives a detection rate of 29\% (including
unconfirmed and not associated with IRAS position sources) or 21\%
if only the confirmed IRAS associated sources are considered.
This is very similar to the 22\% detection rate of Cohen et al.
\cite{cohen88} although the Cohen et al.  sample contained
only sources with F$_{60}>1000$ Jy. Cohen et al.  also found that
higher F$_{60}$ was correlated with a higher probability of the
presence of OH masers.  There is little evidence for such a
correlation in the objects observed in this survey, although the
small number of high flux sources in this sample makes it
difficult draw any firm conclusion.

\begin{figure*}
\begin{center}
\includegraphics[angle=0,width=15cm]{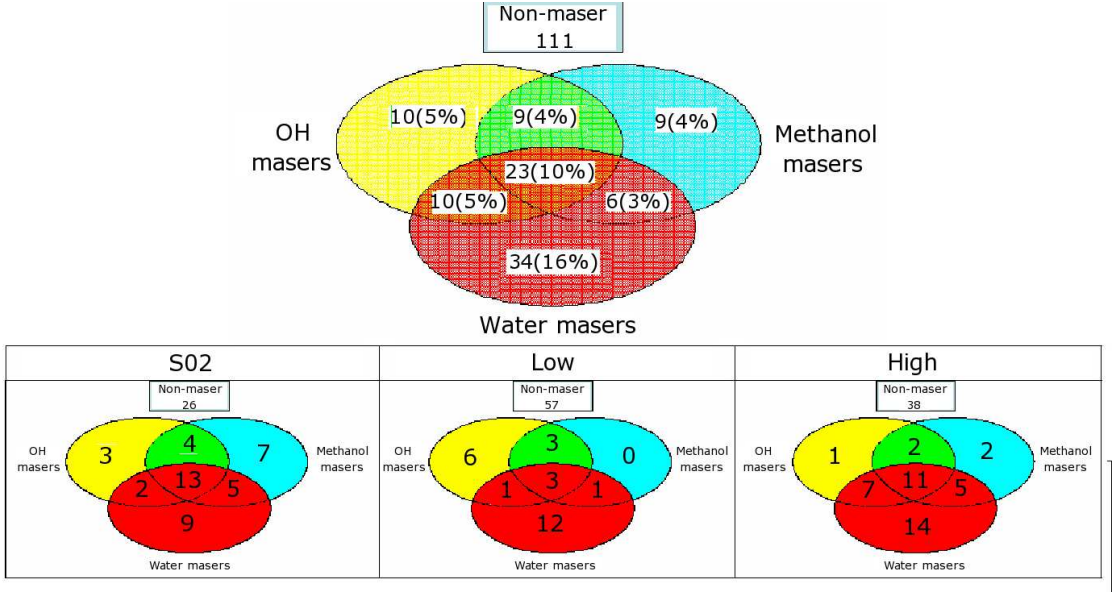}
\caption{Upper panel: Number (and percentage of sample) of
detected sources of the 217 HMPOs sample based on which masers are
present. Note that the 7 offset star forming region candidates in
Table~\ref{tab:OHmaserRes} are considered here while the 5 OH/IR
candidates detected are excluded.
More than half of the sample do not show maser emission. Bottom
panels: same as upper panel but divided by subsamples. For the
whole sample there are 100 sources associated one or more types of
maser, while there are 43, 26 and 42 maser sources in the S02,
\textit{Low} and \textit{High} subsamples respectively. Note that
some sources are common in the S02 and \textit{High} subsamples.}
\label{fig:maser-intersections}
\end{center}
\end{figure*}

Figure~\ref{fig:maser-intersections} summarises the detection
statistics for OH, \htwoo\ and 6.7~GHz, { Class II}, \methanol\ masers
towards the sample. OH and \methanol\ masers show very similar
percentages of detections. Indeed, among the detected OH masers
sources, the two types of maser have 67\% of sources in common. The
number of detections of the three maser types towards the
sub$-$samples is also shown. The S02 and \textit{High} samples show
similarly higher detection rates than the \textit{Low} sample for
which 70\% of its sources have no maser emission. Not only are masers
of any type relatively rare towards the \textit{Low} sub$-$sample,
sources with all three types of maser are particularly uncommon.  In
the \textit{Low} sub$-$sample there are only three such sources,
namely IRAS~19092+0841, IRAS~18024$-$2119 and IRAS~18144$-$1723.
Water masers show the highest detection rates towards the
\textit{High} sample (46\%), while \methanol\ masers show high
detection rate towards S02 sample (42\%).  Perhaps surprisingly there
is a significant difference in the detection rates of S02 and
\textit{High} sample for \methanol\ maser, 42\% and 25\% respectively.
OH masers have similar detection rates towards the \textit{High} and
S02 samples, $\sim$ 26\%. The sources which show only OH maser
emission are mostly from the \textit{Low} sample.  This point is
discussed in \S~\ref{sec:discussion1}.

\subsection{OH Maser Flux Densities}

\begin{figure}
\resizebox{\hsize}{!}{ \includegraphics[angle=-90]{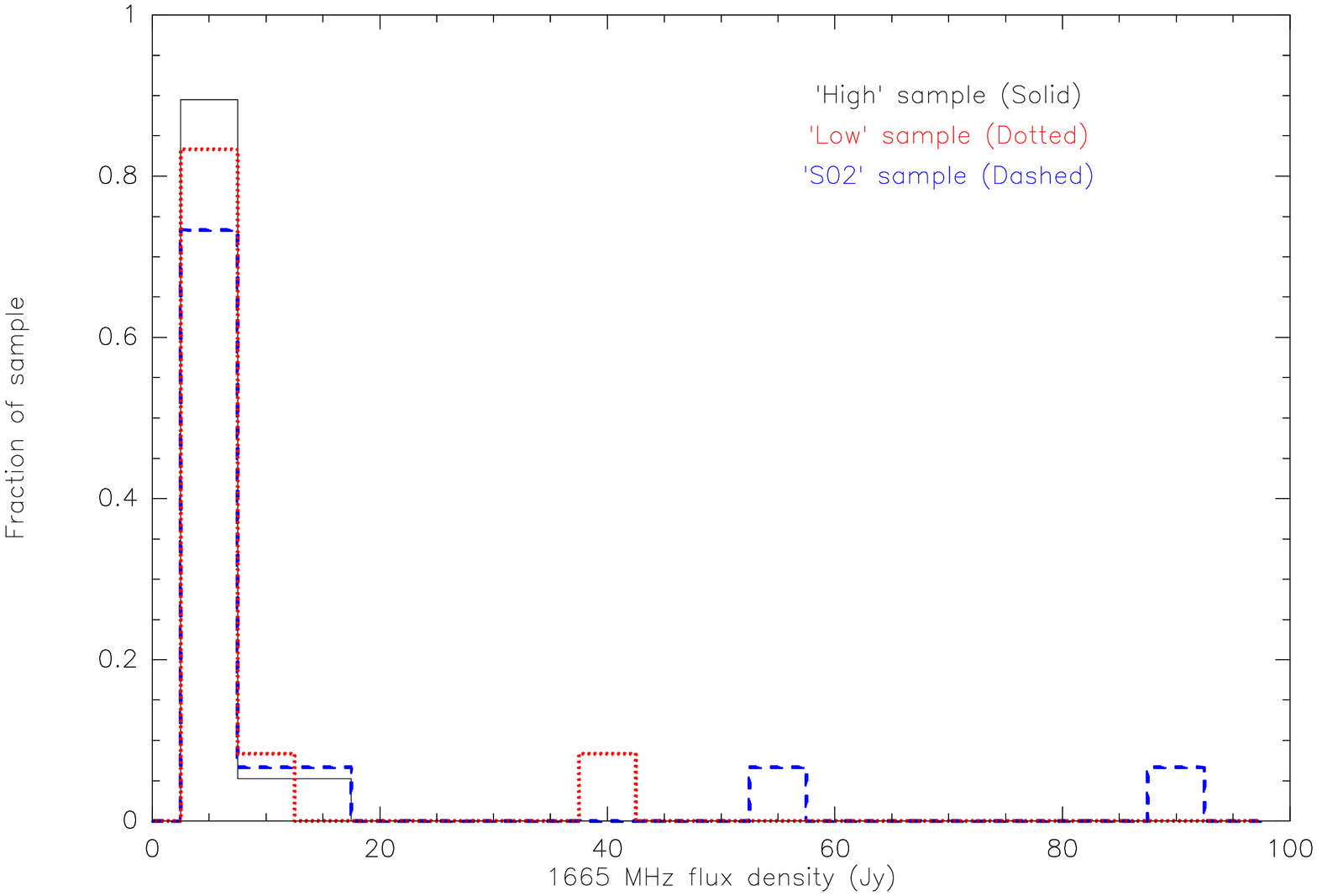}  \includegraphics[angle=-90]{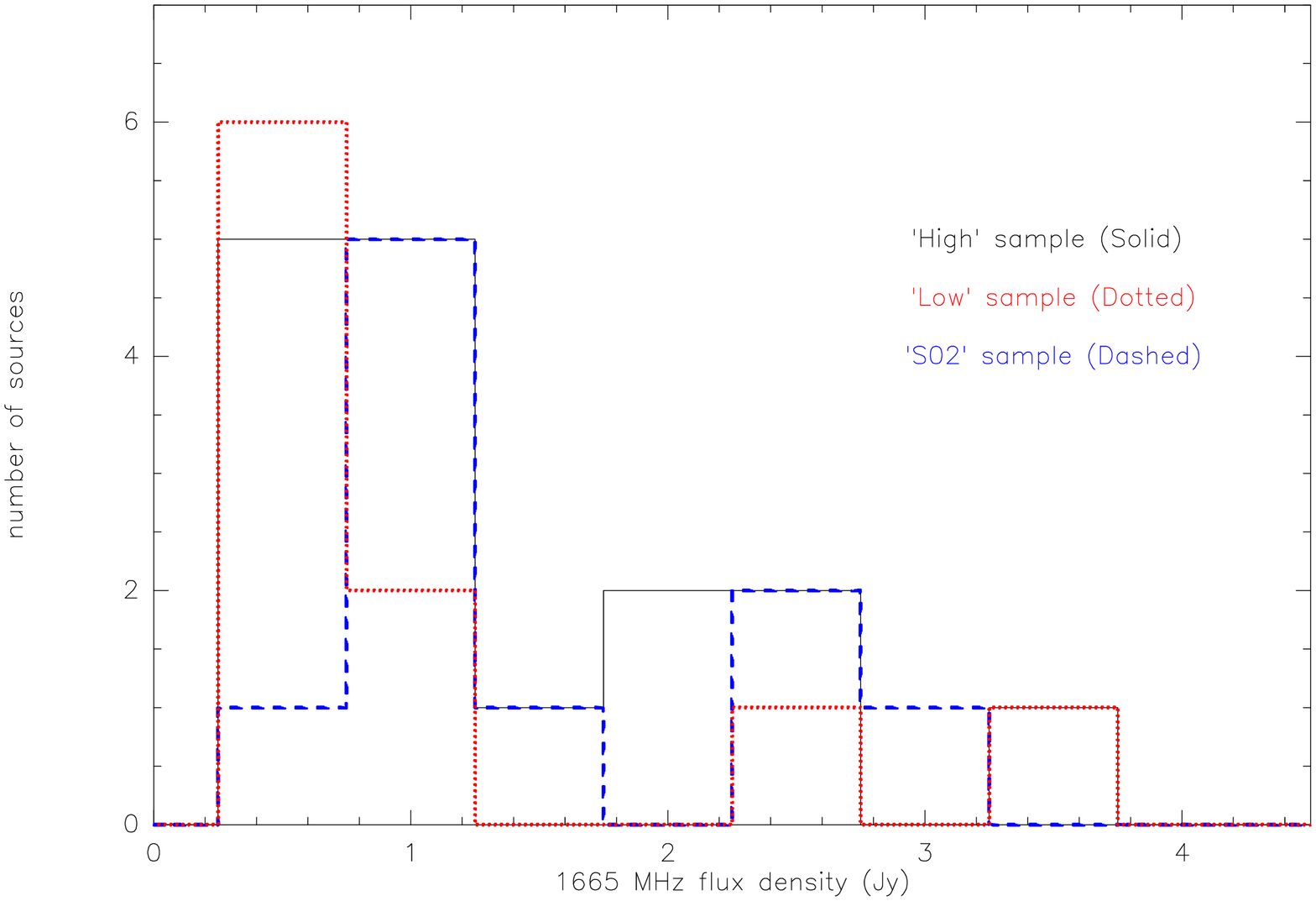}}
\caption{The flux densities at 1665 MHz line towards the three
sub$-$samples, \textit{High} sample (solid line), \textit{Low}
sample (dashed line) and S02 sample (dotted line). An expanded
view for the 1665 MHz flux densities weaker than 4.5~Jy is
re-plotted in the righthand panel.}
 \label{fhsl}
\end{figure}

Figure~\ref{fhsl} shows the distribution of 1665~MHz peak flux densities
towards the three sub$-$samples, \textit{High} sample (solid line),
\textit{Low} sample (dashed line) and S02 sample (dotted line). The right
panel of Figure~\ref{fhsl} shows an expanded view for sources with 1665 MHz
flux density weaker than 4.5~Jy. These figures show that nearly half of the
detected sources have peak flux densities $\lesssim$ 1 Jy. The significance of
this result is discussed in Sec.~\ref{sec:discussion5}.  Figure~\ref{fhsl}
also shows a further difference between the different sub$-$samples.  Most of
the \textit{Low} sample have 1665 MHz flux densities  between 0.25 and 0.75
Jy while most of those of the S02 sample between 0.75 and 1.25 Jy and the
\textit{High} sample shows a wider spread of 1665 MHz flux densities.

\subsection{IRAS flux densities and colour analysis}
\label{sec:IRAS fluxes and color analysis}

\begin{figure}
\centering
\resizebox{\hsize}{!}{\includegraphics[angle=-90]{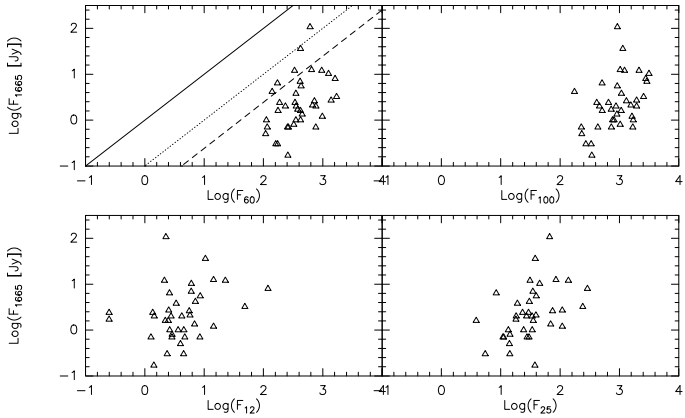}}
 \caption{OH 1665 MHz peak
flux densities of the detected masers are plotted against IRAS
flux densities at 12, 25, 60 and 100~$\mu$m. The solid line, in
the top-left panel, corresponds to equal flux densities, the
dashed line is the line of Slysh et al. \cite{Slysh97} data
(corresponding to $F_{OH}$ = 0.024~$F_{60}$) and the dotted line
is the line of Moore et al. \cite{moore88} data (corresponding to
$F_{OH}$ = 0.1~$F_{60}$).}
 \label{fig:F_IR-65}
\end{figure}

The correlation of maser emission and IR has been studied by many authors to
examine if masers are pumped by IR photons or not. A correlation of OH maser
flux and IR flux has been clearly seen at 60 and 100$\mu$m by several authors
(Cohen et al \citealp{cohen88}; Moore et al.  \citealp{moore88}; Slysh et al.
\citealp{Slysh97}). Figure~\ref{fig:F_IR-65} plots the 1665~MHz peak flux
densities against the IRAS flux densities at 12, 25, 60 and 100~$\mu$m for the
sample observed here. The top-left panel of F$_{60}$ $vs$ F$_{1665}$ also
shows the results of previous studies: the Slysh et al. \cite{Slysh97} data
upper limit (dashed line) corresponds to $F_{OH}$ = 0.024 $F_{60}$ and the
dotted line is the upper limit line of Moore et al. \cite{moore88} data
(corresponding to $F_{OH}$ = 0.1 $F_{60}$). The solid line corresponds to
equal flux densities.  Although the sources searched here have flux densities
similar to Slysh et al.  \cite{Slysh97}, the survey results are more
consistent with the Moore et al. \cite{moore88} line and appear to be
consistent with an extension of Moore et al. sample to lower IR flux
densities. The distribution at 60 and 100$\mu$m flux densities confirm
previous conclusions that, at these wavelengths, a minimum IR flux density is
required for a given maser line strength (Cohen et al.  \citealp{cohen88};
Moore et al. \citealp{moore88}; Slysh et al.  \citealp{Slysh97}).

\begin{figure}
\centering
\resizebox{\hsize}{!}{   \includegraphics[angle=-90]{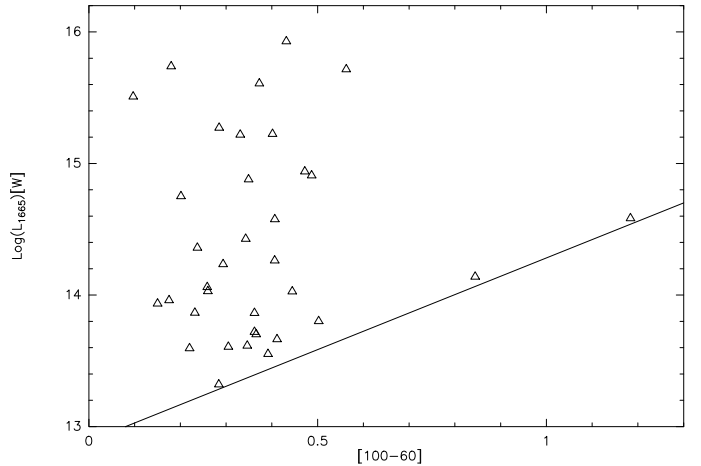}}
  \caption{The 1665 MHz OH luminosities of the detected masers are
    plotted against the [100-60] colour index. There may be a
    suggestion of a lower envelope to the distribution at about
    Log$L_{1665}\sim 0.7 \times$ [100-60] (solid line). Three sources
    with suspect distances (IRAS18024$-$2119, 20062+3550 and
    20227+4154) are not shown on the plot.}
 \label{fig:L65-[100-60]}
\end{figure}

Figure~\ref{fig:L65-[100-60]} plots the [100$-$60] colour index $vs$
OH 1665~MHz maser luminosity, $L_{1665}$. There may be a suggestion of
a lower envelope to the distribution at $\log L_{1665}\sim
0.7$~[100-60] indicating masers are found prefentially towards sources
with warmer radiation temperatures.

\begin{figure*}
\begin{center}
\resizebox{\hsize}{!}{
 \includegraphics[angle=-90,width=8cm]{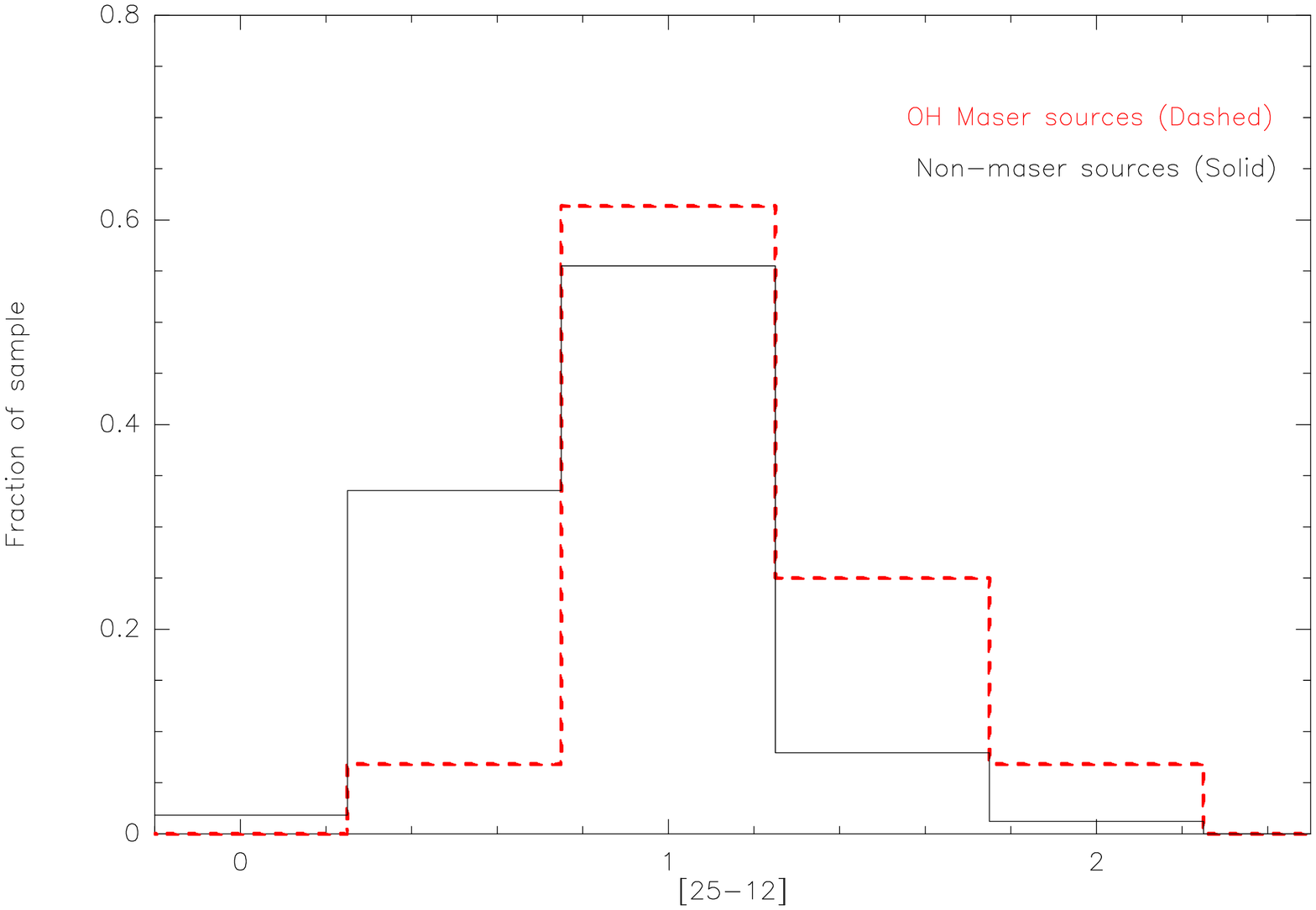}
  \includegraphics[angle=-90,width=8cm]{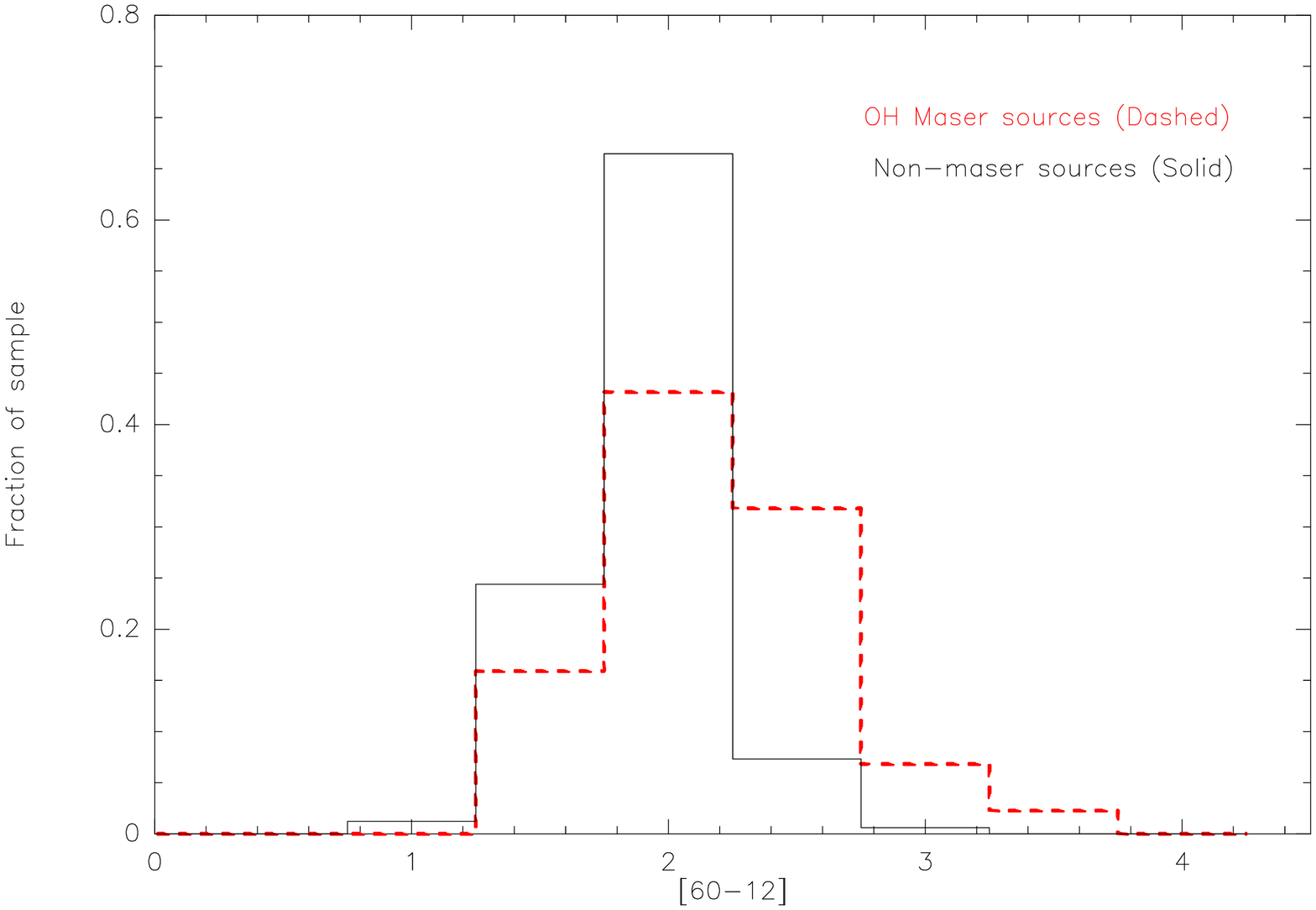}
 \includegraphics[angle=-90,width=8cm]{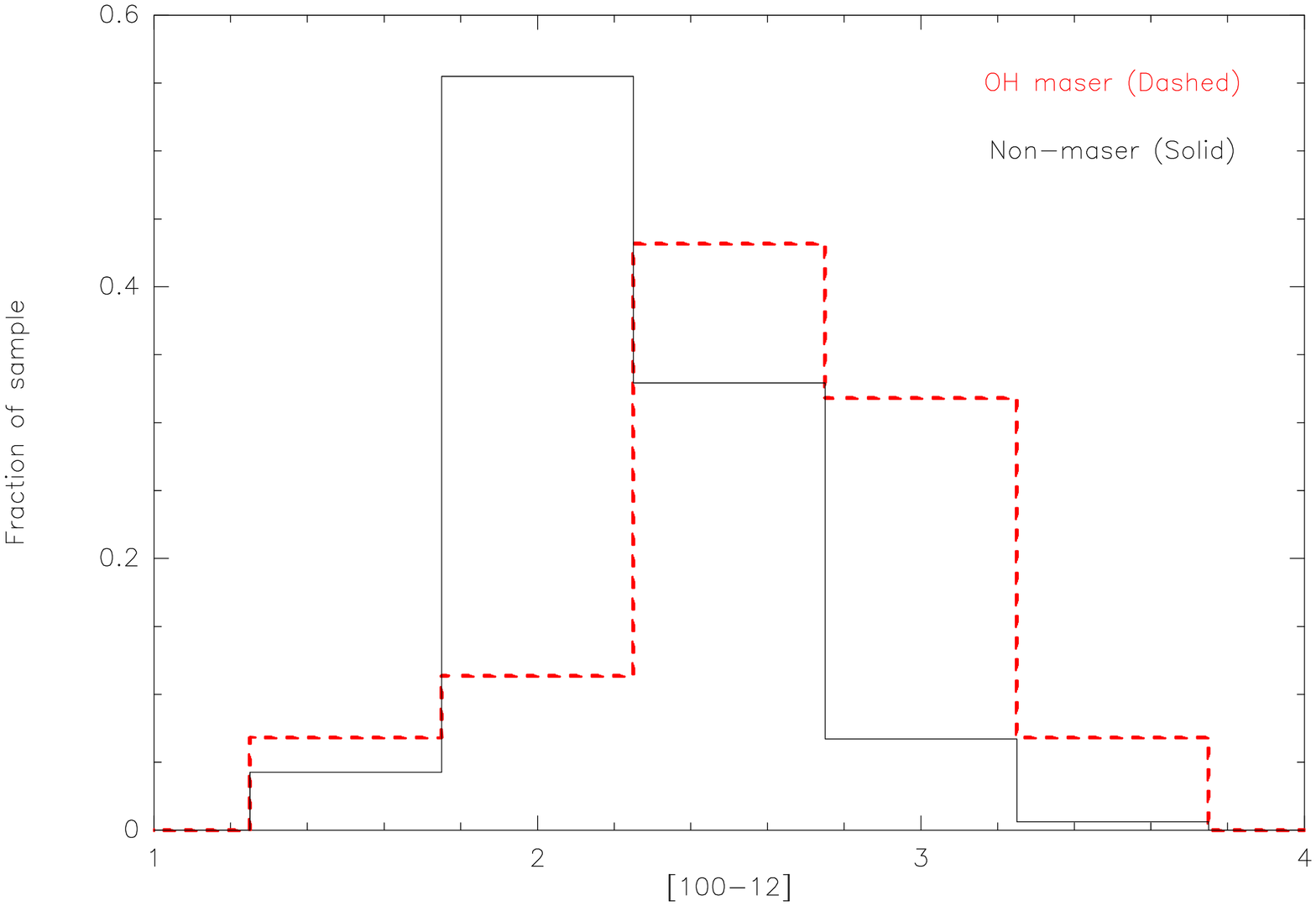}
}
\caption{The [25$-$12] (left panel), [60$-$12] (middle panel) and
  [100$-$12] (right panel) colour indices for OH maser sources (dashed
  lines) and non$-$OH-maser sources (solid lines). In each case OH
  maser sources have on average redder colours than the non-maser
  sources and objects with colours [25$-$12]$>$~1.2,
  [60$-$12]$>$~2.2 and [100$-$12]$>$~2.7 are dominated by sources
  with OH masers.}
 \label{[25_60-12]m-n}
\end{center}
\end{figure*}

\begin{figure*}
 \begin{center}
\resizebox{\hsize}{!}{   \includegraphics[angle=-90]{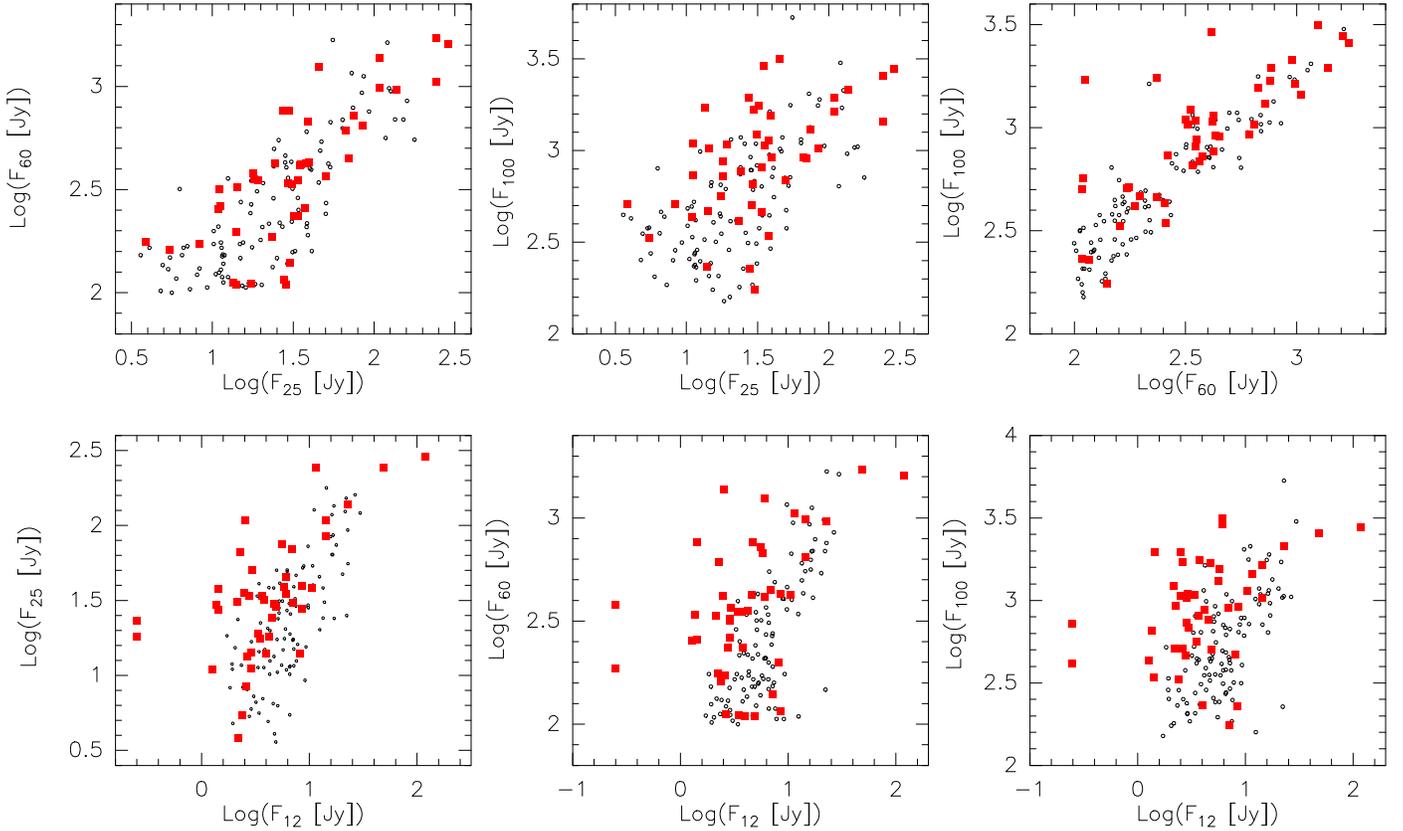}}
  \caption{Comparison of the IRAS flux densities for sources with OH
    masers (red squares) and those without OH masers (black dots).
    The difference between flux densities of maser source and
    non-maser ones is apparent in the lower panels where the effect of
    the 12$\mu$m is obvious, while there is no apparent difference in
    the upper panels.}
 \label{fig:IR-IR-m-nm}
 \end{center}
\end{figure*}

\begin{table*}
  \begin{center}
\begin{tabular}{c c c c c c c c c c c}
  \toprule
  \toprule
                             &[25$-$12]&[60$-$12]&[60$-$25]&[100$-$60]&[100$-$12] & &  $F_{12}$ & $F_{25}$ & $F_{60}$ & $F_{100}$ \\
Maser type/s                 &   \%    &   \%    &   \%    &   \%     &    \%     & &  \%       &   \%     &   \%     &   \%      \\
\midrule		                                                            						    
OH                           &  96.5   &   100   &   53.6  &  19.7    &  100      & &    98.8   &  84.8    &  92.4    &   96.5    \\
H$_{2}$O                     & 100     &   100   &   61.3  &  12.4    &  100      & &    83.1   &  99.2    &  99.7    &   99.5    \\
CH$_{3}$OH                   &  99.9   &   100   &   91.7  &  98      &  100      & &   100     &  41.1    &  99.6    &  100      \\
OH, H$_{2}$O \& CH$_{3}$OH   &  96.5   &   100   &   81.5  &   74.5   &  100      & &    99.9   &  81.5    &  99.4    &   99.4    \\
\bottomrule
\end{tabular}
\caption{Results of KS test for the maser and non-maser population
being drawn from the same populations. The table shows the
probability \textit{D} that the maser and non-maser sources are
drawn from different populations. Clearly the most consistent
difference is in [60$-$12] and [100$-$12] for which maser sources
of all types have consistently different colours than the
non-maser sources. Values greater than 99.9\% shown as 100\%.
For comparison a $3\sigma$ result would correspond to \textit{D} =
99.87\%.} \label{p}
\end{center}
\end{table*}

To see if there is a difference in IR flux densities or colours
between maser and non-maser sources, IRAS sources associated with one
or more type of maser were compared with the rest of the sources.
Table~\ref{p} gives the results of Kolmogorov$-$Smirnov (KS) test for
the probability \textit{D}, that the flux density or colour
distribution between maser and non$-$maser source are different. The
table shows that in general, the maser sources do show different flux
densities and colours from non$-$maser ones.  The difference between
maser sources and non$-$maser sources is most evident in the [60$-$12]
and [100$-$12] colours for all types of masers where \textit{D}
$\equiv$ 100\% and in the [25$-$12] colour for \htwoo\ and \methanol\
masers where \textit{D} $\geq$ 99.9\%. Figure~\ref{[25_60-12]m-n}
plots the distribution of these three colour indices { for the OH }
maser and non-maser sources to demonstrate the last points.  { It
  can also be seen in Figure~\ref{fig:[60-25-12]}, where it is clear
  that sources with OH masers dominate over non-maser sources in the
  regions [25-12]$>1.2$ and [60-12]$>2.2$. Even for colours $\sim0.2$
  smaller than these values, the maser sources make up $\sim50$\% of
  the population, a higher fractor than for sources with smaller
  colours and for the sample overall. The redder colours of the maser
  sources is unlikely to be affected by the uncertainties in the
  colours of the sources unless the uncertainties preferentially
  affect maser sources and non-maser sources in different senses.}

The [25$-$12] and [60$-$12] colour indices are the ones used by WC89
to identify UCHII regions. These results show that maser sources tend
to be relatively faint at 12$\mu$m compared to their fluxes at longer
wavelengths.  Figure~\ref{fig:IR-IR-m-nm} shows the four IRAS flux
densities against each other for maser sources (squares) and non-maser
sources (dots).
The difference between maser sources and non-maser sources is obvious
in the lower panels which include the 12$\mu$m flux densities.  On the
other hand, as Table~\ref{p} also shows, sources with only OH or
H$_2$O masers have [60$-$25] and [100$-$60] colours indistinguishable
from non-maser sources.

\begin{figure}
\resizebox{\hsize}{!}{
 \includegraphics[angle=-90]{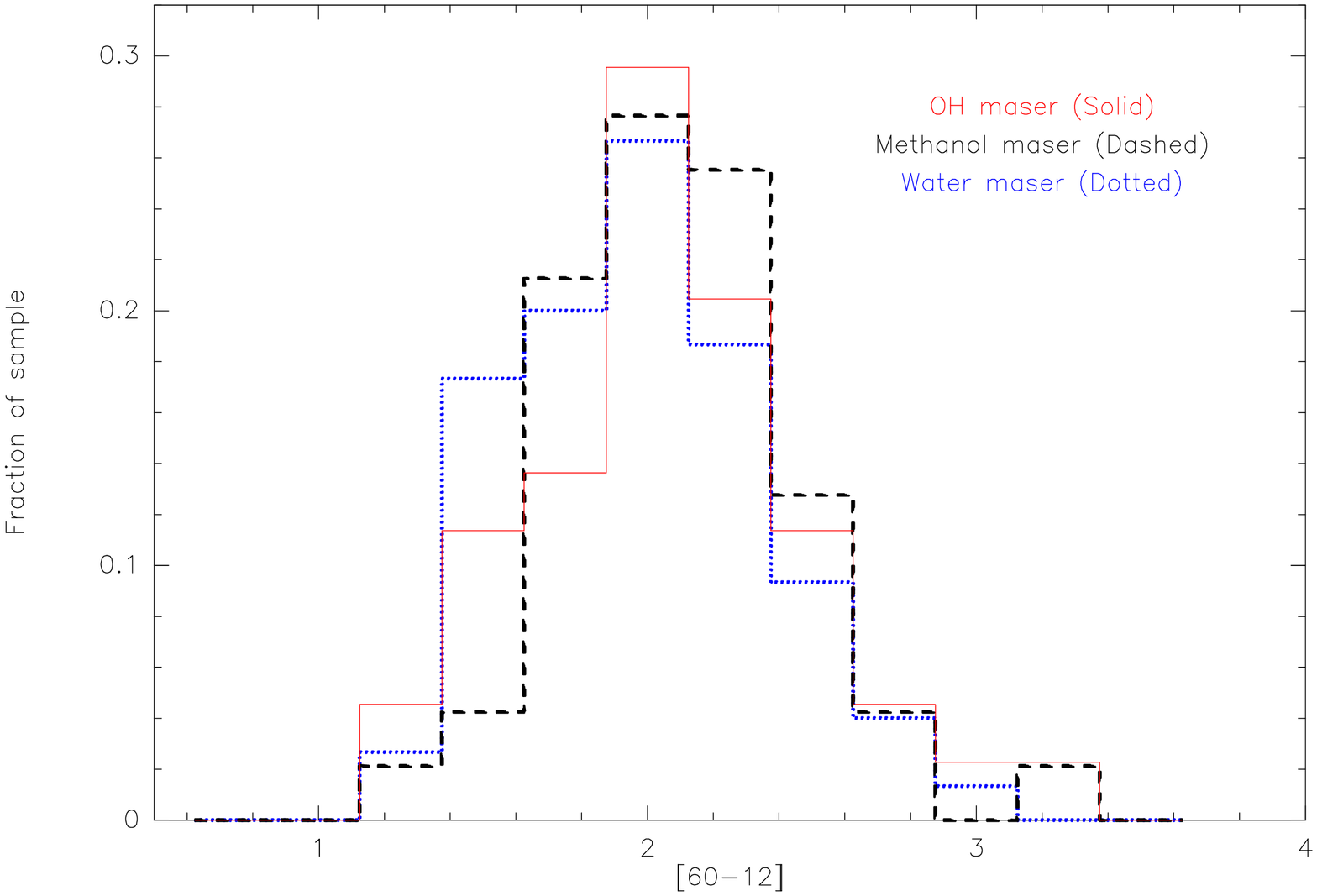}
  \includegraphics[angle=-90]{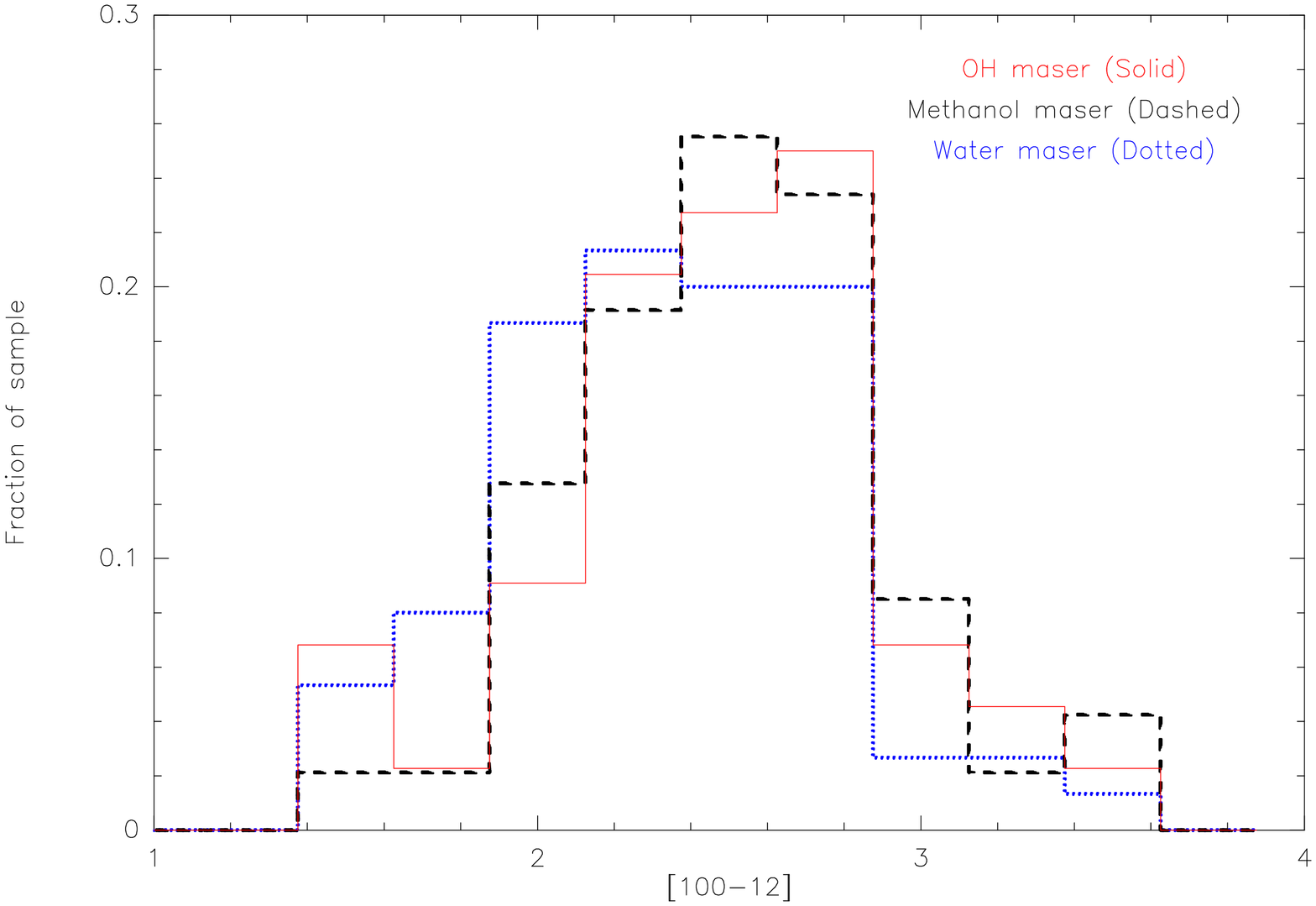}
}
 \caption{[60$-$12] and [100$-$12] colours of sources with any
 type of maser, water (dotted lines), methanol (dashed lines)
 and OH (solid lines). There is no significant difference in
 the distribution of these colours as a function of maser type.}
 \label{fig:colcomp}
\end{figure}

Given the differences in [60$-$12] and [100$-$12] colours of sources with any
type of maser, it is interesting to ask whether there is any systematic
difference in source colour as a function of maser type.
Figure~\ref{fig:colcomp} compares these colours for sources with each type of
maser showing that there is no significant difference in the distribution of
these colours as a function of maser type.  For this sample there appears to
be no evidence in their IRAS colours that different types of maser trace
sources in significantly different evolutionary states.

\begin{figure}
\centering
\resizebox{\hsize}{!}{   \includegraphics[angle=-90]{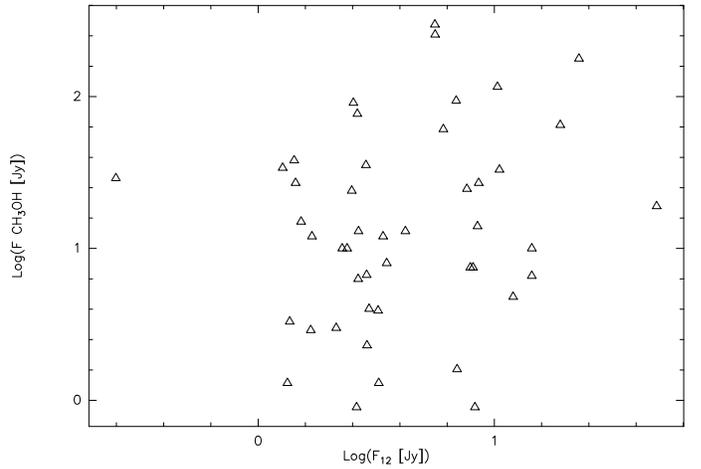}}
  \caption{Methanol maser peak flux densities are plotted against IRAS
    flux densities at 12~$\mu$. No correlation is apparent.}
 \label{fig:f_12-f_ch3oh}
\end{figure}

Regarding the 100, 60, 25 and 12$\mu$m flux densities, the probabilities,
\textit{D} in Table~\ref{p} are relatively large for all flux densities,
except the \methanol\ only sources at 25$\mu$m. However only for few
maser--flux combinations are the individual \textit{D} probabilities
statistically significant at a 3$\sigma$ level. Although methanol masers are
believed to be excited by mid-IR photons (Cragg et al. \citealp{Cragg05}),
Figure~\ref{fig:f_12-f_ch3oh} shows that there is no clear correlation between
\methanol\ maser and 12~$\mu$m flux densities.

\begin{figure}
\centering
\resizebox{\hsize}{!}{    \includegraphics[angle=-90]{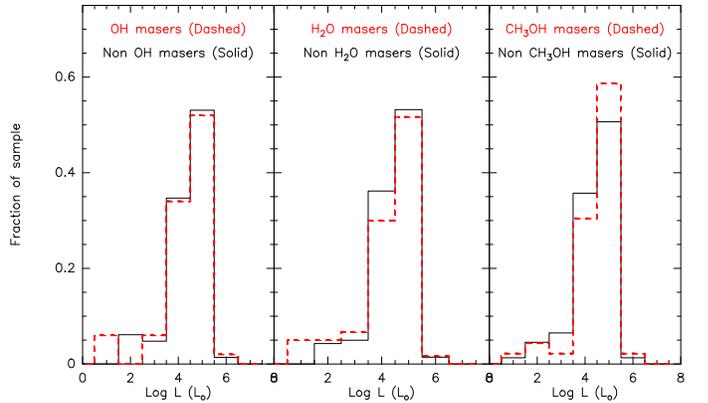}}
\caption{Comparison of the luminosity of maser sources and
non-maser sources. Note that the panels show sources with a
specific maser type with sources are not associated with this type
but may be associated with other types of masers.}
 \label{fig:Lum-masers}
\end{figure}

Figure~\ref{fig:Lum-masers} compares the luminosity of maser sources and
non-maser sources. The distributions are statistically indistinguishable,
indicating that masers are not preferentially associated with only the more
luminous sources. 

\begin{figure}
  \centering
\resizebox{\hsize}{!}{  \includegraphics[angle=-90]{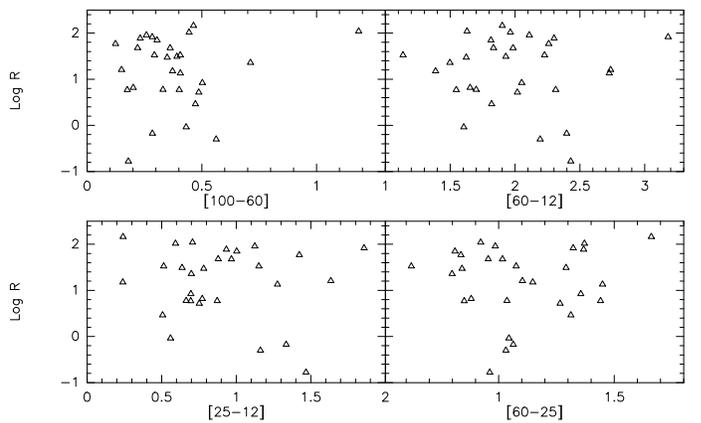}}
\caption{Ratio of the 6668 MHz \methanol\ peak flux density to the
1665 MHz OH peak flux density, log\textit{R}, versus IRAS
colours.}
 \label{fig:colors-67-65}
\end{figure}

Comparing the IRAS colour indices to the ratio of \methanol\ and OH maser flux
densities, Figure~\ref{fig:colors-67-65} plots the ratio of the OH peak flux
density to the \methanol\ peak flux density, \textit{R}, versus the IRAS
colours of a source.  According to Caswell \cite{caswell98} the spread in the
ratio \textit{R} reflects the range in evolutionary stage of the sources, but
this figure does not show any evidence that \textit{R} depends on the colour
of a source which might also be expected to evolve as a source evolves. In
considering this result, it should be noted that the IRAS colours may suffer
from confusion because of the poor spatial resolution of IRAS and maser
observations combined with the tendency of massive stars to form in clusters
(cf. Bourke et al. \citealp{bourke05}).

\subsection{Velocity Range of Masers}

\begin{figure}
\resizebox{\hsize}{!}{ \includegraphics[angle=-90]{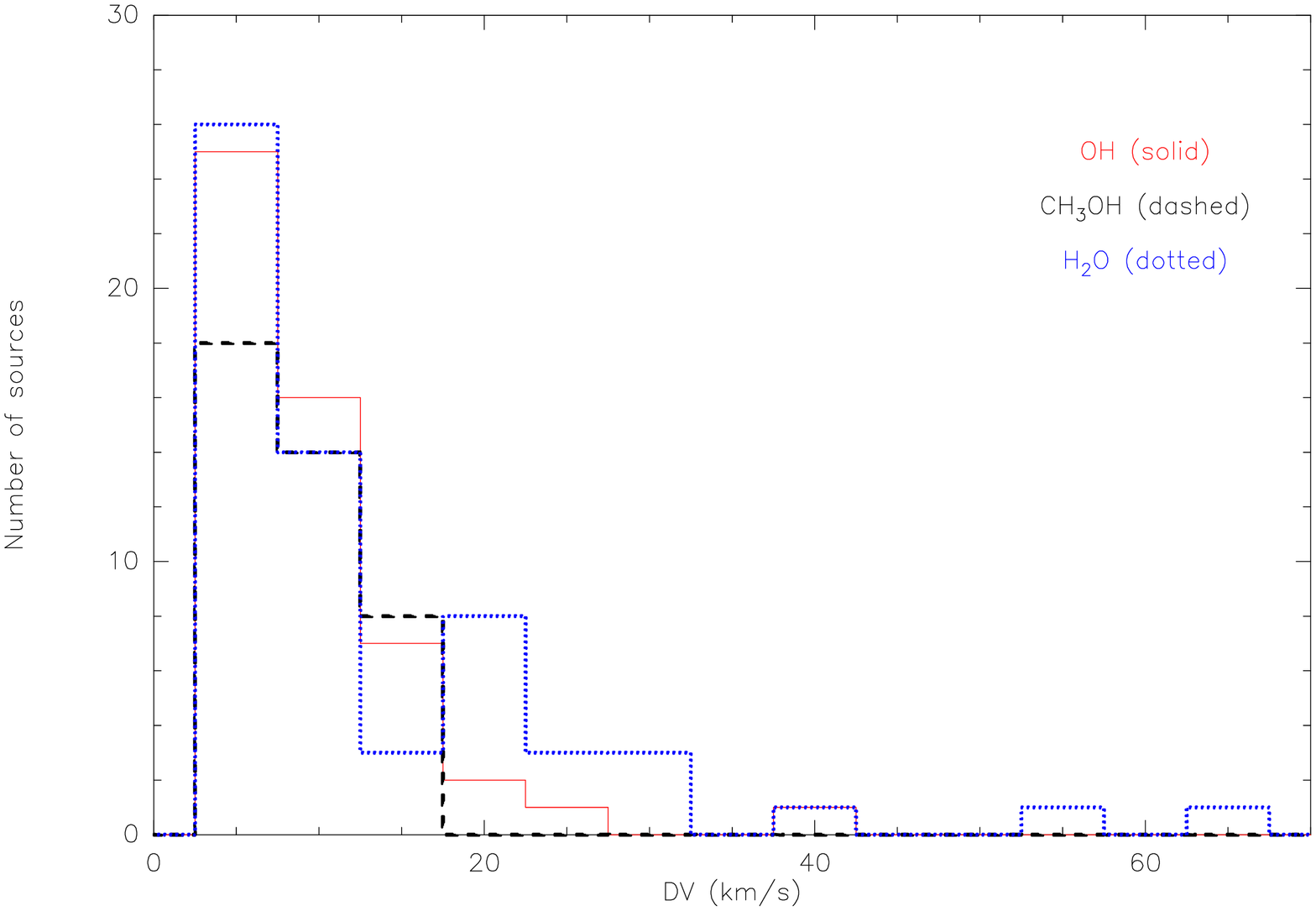}  \includegraphics[angle=-90]{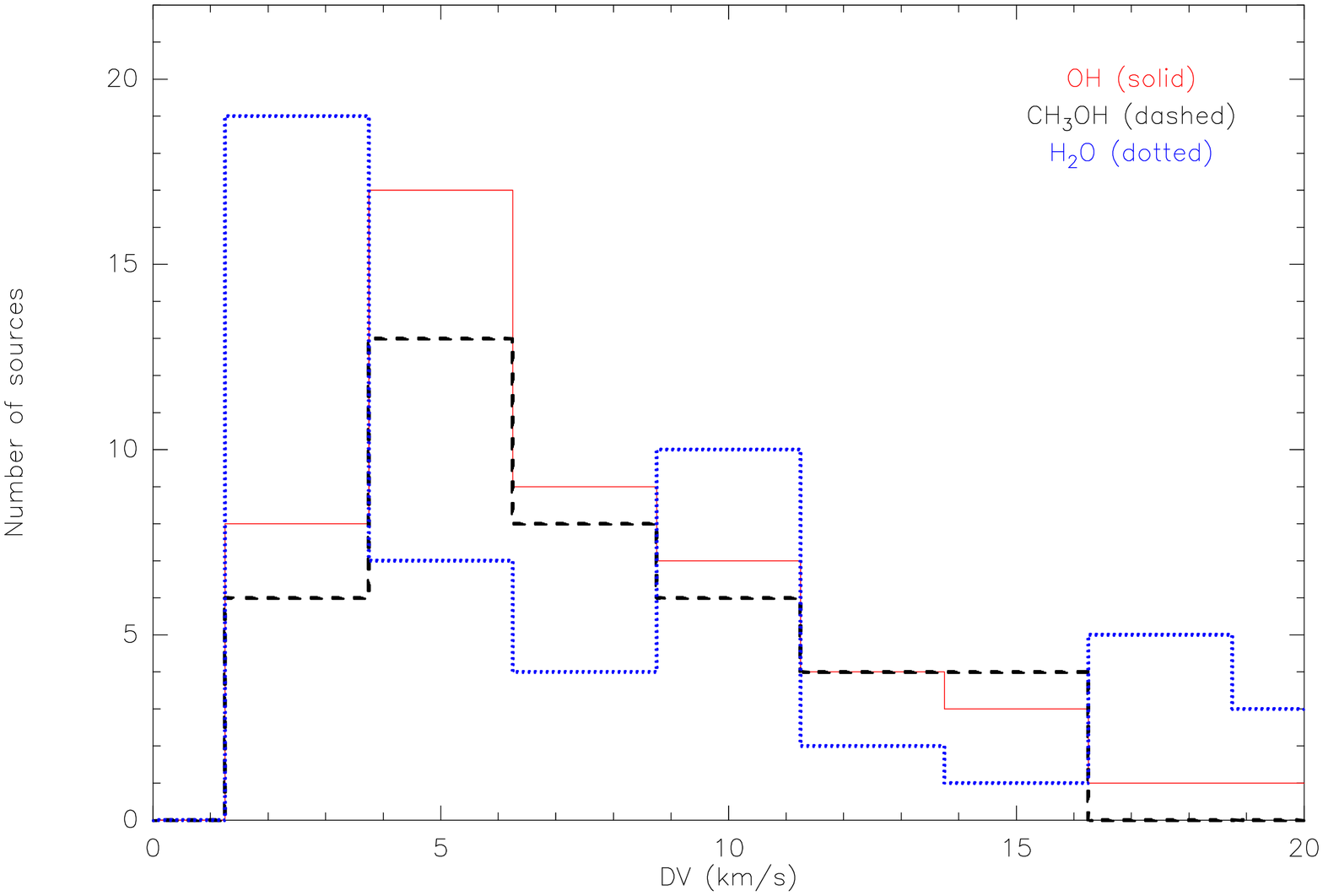}}
  \caption{The velocity spread of the masers associated with the whole
    sample: OH (solid line), \methanol\ (dashed line) and \htwoo\
    (dotted line). The righthand panel is an expanded view for velocities
    spreads $\leq$ 20~km \s.}
 \label{fig:DV-all}
\end{figure}

The range of velocities over which maser emission is observed can be
used to investigate whether the emission from different masers arises
from different material associated with the sources.
Figure~\ref{fig:DV-all} shows the distribution of the range of
velocity for each OH maser source detected
(Table~\ref{tab:OHmaserRes}). For comparison the figure also shows the
velocity range of the \htwoo\ and \methanol\ masers from SHK2000, P91,
S02 (and references therein) and Han et al.  (1998). The righthand
panel of Figure~\ref{fig:DV-all} shows an expanded view of the sources
with velocity range $\leq$ 20~km \s.

For all three species the velocity range peaks at less than 10 km~\s. The
CH$_{3}$OH masers and OH masers extend up to velocity ranges of 17.5 and 22.5
km~\s\ respectively.  Overall the figures show that the OH and \methanol\
agree very well in their distribution of velocity range, peaking at $\sim5$
km~\s, suggesting that these masers may originate in similar material around
the sources.  On the other hand the \htwoo\ masers show a quite different
distribution. In some sources the emission covering up 65 km~\s\ has been
observed. However the distribution actually peaks at spreads $<4$ km~\s,
smaller than for the OH and \methanol. This suggests a two population
structure, with one group of sources with velocity ranges between $\sim1$
km~\s\ and $\sim4 - 5$ km~\s\ and a second with velocity ranges $>5$ km~\s.

\begin{figure*}
\resizebox{\hsize}{!}{
 \includegraphics[angle=-90,width=0.66\hsize]{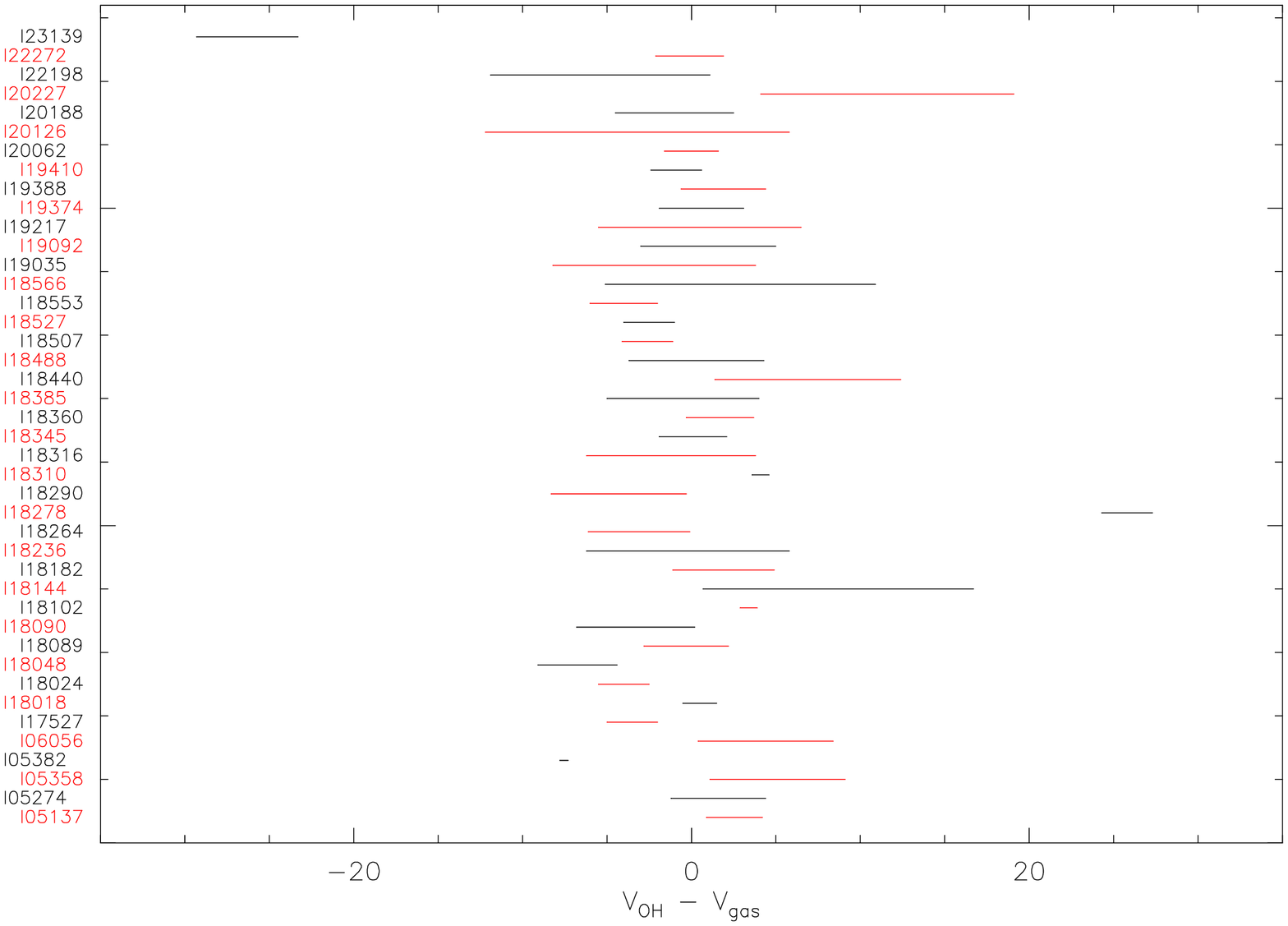}
  \includegraphics[angle=-90,width=0.66\hsize]{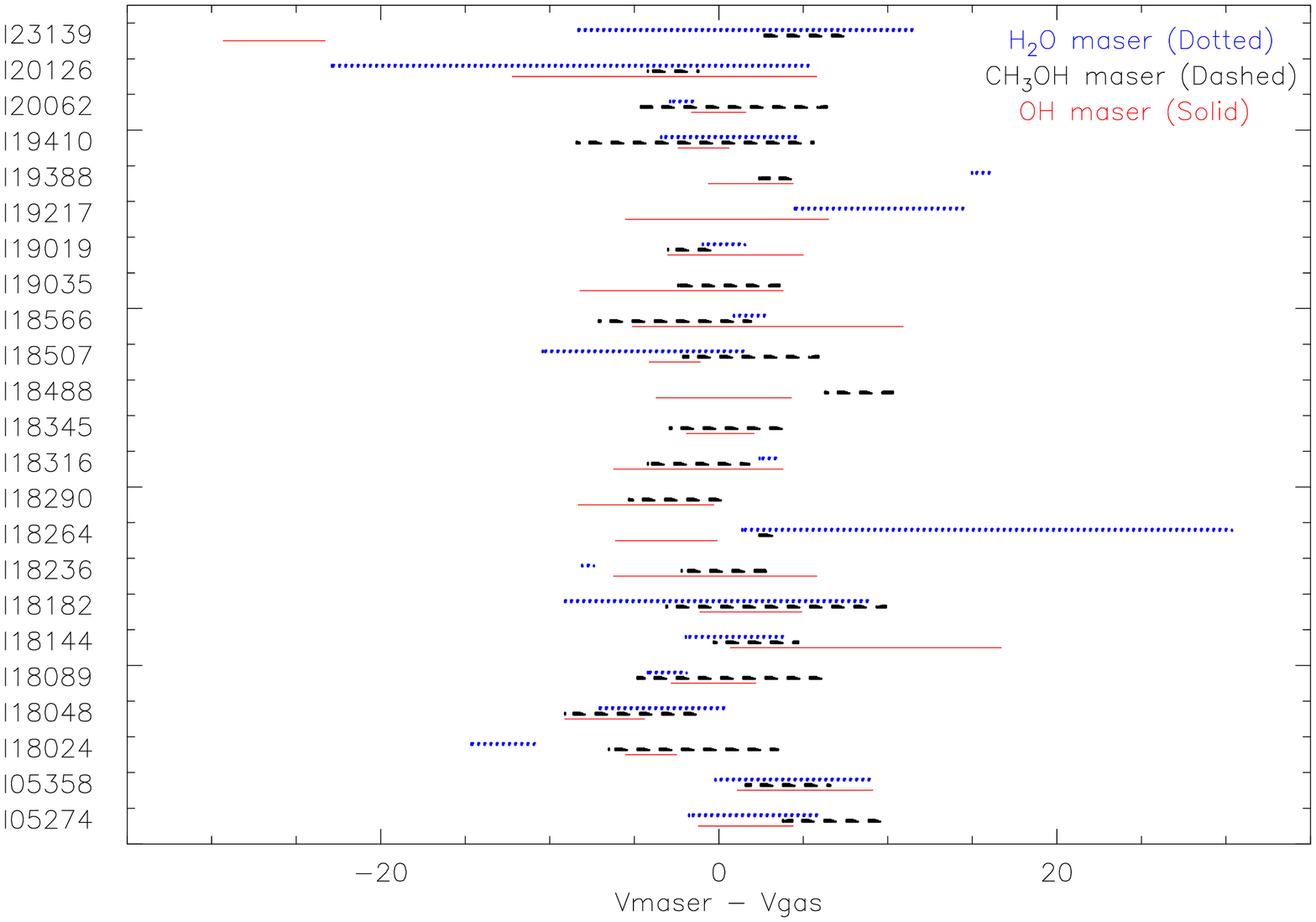}
}
\caption{\textit{Left hand panel}: the velocity range relative to
  cloud gas velocity for OH maser sources. \textit{Right hand panel}:
  the same as left hand panel but for the 23 sources showing three
  maser types. Note that some sources, although detected in \htwoo\
  and/or \methanol, do not have published information on their
  velocity extent.}
 \label{fig:Vmaser-Vgas}
\end{figure*}

The offset between the OH maser velocity and the velocity of the
dense gas towards each source (M96 and S02) is shown in
Figure~\ref{fig:Vmaser-Vgas} (left). Overall the OH maser
velocities are distributed around the gas velocity, as they are
also for some particular sources. On the other hand there are
exceptions with some sources where the OH masers are offset by up
to $\sim20$km~\s\ from the gas velocity.  Comparing the velocity
offset between different maser type and dense gas for sources with
all three types of masers (Figure~\ref{fig:Vmaser-Vgas}, right),
it is difficult to identify any global trend. However inspection
of the observations shows 8 out of the 23 sources with all three masers
have $\Delta$V(OH) $>$ $\Delta$V(\methanol) $>$ $\Delta$V(\htwoo).

\section{Discussion}
\label{sec:discussion}

\subsection{Comparison of OH Masers With 6668 MHz Methanol Masers}

The similarity in detection rates and velocity ranges suggest an
association between OH and { Class II} \methanol\ masers in these sources
as first suggested by Caswell et al.  \cite{caswell95} and modelled by
Cragg, Sobolev \& Godfrey \cite{cragg02}.  A similar result has also
been found by  Szymczak \& G\'{e}rard \cite{szymczak04} who searched
a sample of 100  \methanol\ maser sources for OH masers. They
found that 55\% of \methanol\ maser sources also have OH maser
emission.  Their results also show that OH and \methanol\ masers cover
similar velocity ranges.

\begin{figure}
\resizebox{\hsize}{!}{ \includegraphics[angle=-90]{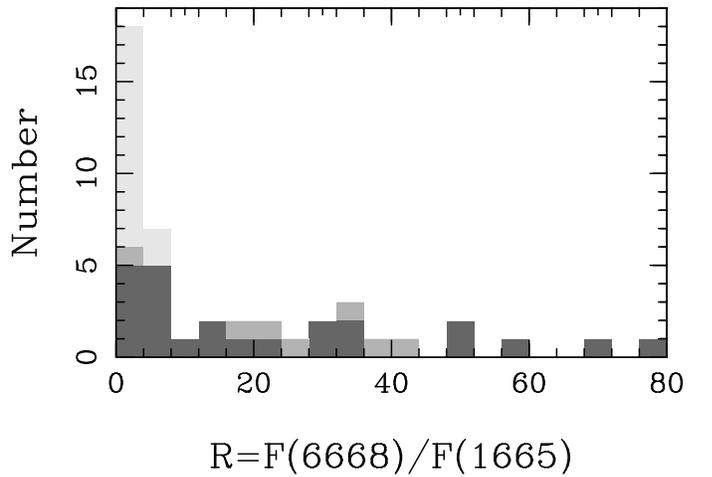}}
\caption{ Distribution of number of sources as a function of $R=
  \frac{F(6668)}{F(1665)}$, the ratio of the peak \methanol\ maser
  intensity to the peak OH maser intensity. The darkest bars show
  measured values. The lightest bars show upper limits on $R$ for
  sources with no detected methanol masers and the mid-grey bars show
  the lower limits on $R$ for sources with no detected OH masers.
  There are 5 sources with measured values of $R>82$ and 5 lower
  limits with $R>82$ which are not shown on the plot. Caswell
  \cite{caswell97} identified sources with $R\le8$ as OH-favoured and
  those with $R\ge32$ as methanol-favoured.}
 \label{fig:1665-6668}
\end{figure}

We applied the suggestion of Caswell \cite{caswell96} for
characterizing the maser sites where methanol and OH masers occur
using the ratio of peak methanol intensity to peak 1665$-$ MHz OH
intensity. Figure~\ref{fig:1665-6668} shows the distribution of the OH
to \methanol\ intensity ratio, $R =
\frac{F(6668)}{F(1665)}$ (Table~\ref{tab:OHmaserRes}).
For the 29 sources which show both 1665~MHz OH masers and 6668 MHz
masers, 12 sources have values of \textit{R} $>$ 32 which places them
in methanol-favoured region and 10 sources in OH-favoured region
(\textit{R} $\le$ 8).

There are an additional 12 
sources which do not have OH maser emission but do have methanol
masers and 20 sources with OH masers but no methanol masers.  Limits
on $R$ for these sources were obtained by adopting a three sigma limit
on the presence of OH masers of 150~mJy (typical of this survey) and a
limit of 1500~mJy for the presence of methanol masers (Pestalozzi et
al. 2005).  The resulting 3-sigma limits are also shown on
Figure~\ref{fig:1665-6668}.  All the methanol non-detections are
consistent with the sources being OH favoured. For the OH
non-detections, one source with a lower limit on \textit{R} of 3 could
be OH favoured, 8 sources are definitely methanol favoured and the
remaining 3 have lower limits on $R$ between 8 and 32.

The sources observed here have a very wide range in $R$, with measured
values from ranging from 0.2 to 145 and three lower limits on $R$ in
excess of 400.  The distribution of these values is in marked contrast
to the results of Caswell \cite{caswell96}, who found most sources to
have $R$ in the range 8 to 32, with a typical value of 16.  Over our
whole sample, including the limits, 25 sources (41\%) are OH-favoured.
A similar fraction of the sample, 26 sources (43\%), are \methanol\
favoured, which Caswell \cite{caswell96} suggested are sources in an
earlier evolutionary stage than OH-favoured sources. Only 7 sources
(11\%) have measured values of $R$ between 8 and 32 plus there are a
further 3 lower limits (corresponding to a further 5\% of the sample)
which could place these sources in this range of $R$.  If $R$ does
indeed trace the evolutionary status of sources as Caswell has
suggested, then these results suggest that the sample observed here
represent a particular mix of sources in different evolutionary stages
distinct from that in the sample observed by Caswell. The high
fraction of methanol favoured sources here suggesting that this sample
contains a higher portion of younger objects.

\subsection{OH maser flux densities}
\label{sec:discussion5}

Figure~\ref{fhsl} shows that nearly half of the detected sources show
OH flux densities $\lesssim$ 1 Jy, which indicates two points. First it
demonstrates how less sensitive observations (with detection limits of
$\gtrsim$ 1~Jy, e.g. Cohen et al.  \citealp{cohen88}) could be missing
a significant population of OH maser sources. More importantly it
indicates that the sources observed here, which are believed to be in
an early stage of formation with no observable UCHII regions, have
lower OH flux densities than sources in more evolved objects
associated with HII regions, where typical OH masers have flux
densities $\geq$1~Jy.

Figure~\ref{fig:L65-[100-60]} shows that most of the OH maser sources
prefer warmer radiation fields. This is probably because most of the
FIR transitions which pump the OH maser are at wavelengths between
$\sim$60$\mu$m and 100$\mu$m (Gray, Field \& Doel 1992).  The
luminosity of the 1665~MHz line spans a factor of about 300 across the
sample, but the origin of this scatter is unclear. However it is
likely that individual maser spots are saturated, so for a given
source luminosity (especially in the spectral region $\sim$60$\mu$m
$\leq \lambda \leq$ 100$\mu$m) the maser luminosity may reflect the
number of maser sites or volume of masing gas (Gray priv. comm.).

Models of OH maser emission show different physical conditions give
rise to different combinations of OH maser lines. For example the
Cragg et al. \cite{cragg02} models show that 1665~MHz masers trace a
wider range of conditions that 1667~MHz masers. In particular gas
kinetic temperatures $>75$K, and gas number densities
$\gtrsim10^7$cm$^{-3}$ more strongly quench the emission at 1667~MHz
than the emission at 1665~MHz suggesting that ratio of these lines
fluxes may be a probe of the conditions in the gas, with masers seen
in only the 1665~MHz line tracing warmer, denser gas.

\begin{figure}
\resizebox{\hsize}{!}{
  \includegraphics[angle=-90]{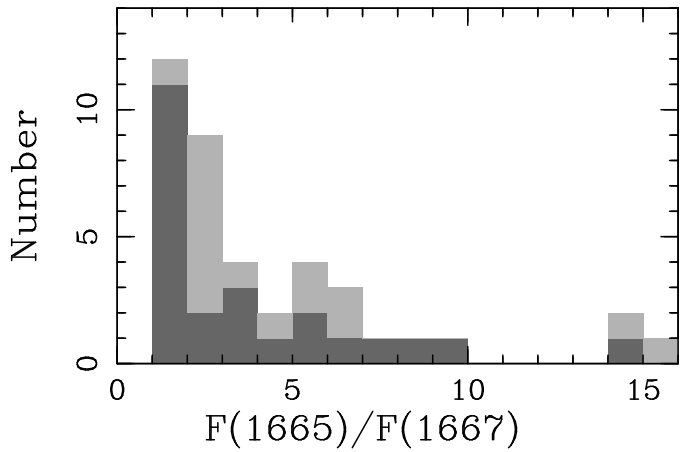}
}
  \caption{ The distribution of the ratio of peak 1665~MHz flux
    density to peak 1667~MHz flux density for the sources. The dark
    coloured bars show the measured ratios.  There are four sources
    for which the 1667~MHz line is stronger than the 1665~MHz which
    are not shown.  The light coloured bars show the 3 sigma lower
    limits for the sources which were detected at 1665~MHz but not
    1667~MHz. There is one source detected at 1667~MHz but not
    1665~MHz which is also not shown. Sources where the masers are
    offset from the IRAS position are included in the distributions.}
 \label{fluxratio}
\end{figure}

For the sample observed here 28 sources were detected in both the
1665~MHz and 1667~MHz lines. For all except four of these objects the
1665~MHz line was stronger that the 1667~MHz line.  The ratio of the
flux density at 1665~MHz to that at 1667~MHz for these objects is
shown in Figure~\ref{fluxratio}.  A further 16 objects were detected
in only the 1665~MHz line. The 3-sigma lower limits on the 1665~MHz to
1667~MHz ratio for these sources are also shown the figure. The
majority of sources have ratios of less $\sim3$, but there is a tail
of sources (and limits) up to ratios as large as $\sim15$. Whether
this range of intensity ratios reflects different physical conditions
in these sources requires more detailed follow-up observations of the
sources, particularly at higher angular resolution.

\subsection{OH Maser Only Sources}
\label{sec:discussion1}

There is a group of newly detected sources associated with only OH
maser emission; namely IRAS 06382+0939, 18408$-$0348, 18540+0220,
18454$-$0158, 18511+0146, 18586+0106, 19118+0945, 19220+1432 and 
20099+3640.
The already known OH maser source IRAS~18018$-$2426 may also be in
this group, although there has been no search for \methanol\ maser
emission towards it.  However for only four of these objects, IRAS
18018$-$2426, 18454$-$0158, 19118+0945 and 19220+1432 are the OH
masers positionally associated with the IRAS source. The OH masers
towards the other sources are offset from the IRAS position and so it
is possible that previous searches for other types of maser may have
missed masers which are also offset from the IRAS position.

Interestingly, six of the ten OH-only objects, including IRAS
18018-2426 and IRAS~19118+0945, are from the \textit{Low} sub$-$sample
of M96 which contains sources which do not have UCHII region IRAS
colours (WC89). However, four of these sources are associated with
radio continuum emission and no equivalent observations has been
carried out towards the other two sources, IRAS~06382+0939 and
19118+0945. 
M96 found a lower incidence of ammonia emission towards their
\textit{Low} sample than the \textit{High} one and suggested that the
\textit{Low} sample contains sources in two different evolutionary
stages. Some sources are in a stage before the existence of an UCHII
region and others are in a stage after the HII region has evolved. The
sources with evolved HII regions were suggested to be a very late HMPO
stage and are expected to be associated with relatively small amounts
of neutral gas.  The absence of detectable ammonia towards
IRAS~18408-0348, 19118+0945 (M96), 18454-0158 and 18540+0220 (S02) but
the presence of only OH masers suggest that these objects could be
relatively evolved.

\subsection{1720 MHz only sources}
\label{sec:1720}

There are five sources, IRAS 04579+4703\footnote{Note that the masers
  towards IRAS 04579+4703 are offset from the IRAS source.},
18264$-$1152, 18553+0414, 19220+1432 and 20188+3928, which have OH
maser emission, but only in the 1720 MHz transition. These sources,
apart from IRAS 19220+1432, are associated with other maser types.
They are, apart from IRAS 04579+4703, also associated with, at most,
weak radio continuum emission ($<$ 1 to 11~mJy). Isolated 1720-MHz
masers are sometimes observed in star-forming regions in association
with 6668-MHz methanol masers (e.g. Etoka, Cohen \& Gray
\citealp{etoka05}), an association which can be understood in terms of
common pumping conditions (Cragg et al. \cite{cragg02}). Other objects
associated with only 1720$-$MHz OH maser emission are believed to be
supernova remnants. Elitzur \cite{Elitzur76} suggests that the level
inversion of these 1720 MHz masers is due to collisional excitation.
Models suggest that the masers originate in the post-shock molecular
gas behind C-type shocks (e.g. Lockett et al. \citealp{Lockett99}).
However, these 4 sources appear to be star-forming regions. If the
Elitzur \cite{Elitzur76} model is correct, the masers may be tracing
post-shock gas towards these sources. Confirmation of this
interpretation will require detailed follow-up of these objects.

\section{Conclusions}

The results of this survey confirm that OH masers are common, not only
towards HII and UCHII regions as previously known, but also towards
less evolved high mass young stars which, although luminous, have not
yet significantly ionised their surroundings.  Indeed towards these
young sources OH masers are as common as the \htwoo\ and { Class II}
\methanol\ masers which have often been been cited as evidence of the
youth of these sources.  Compared to the OH masers associated with
HII/UCHII regions, the OH masers towards these younger sources are
weaker, suggesting that the OH maser flux density could be a crude
evolutionary indicator.

There is no evidence within the observed sample for a luminosity
difference between those sources with OH masers and those without
masers.  However, on average the sources with OH masers have
significantly redder [100-12] and [60-12] IRAS colours than those
which do not have OH masers, suggesting that the OH maser sources are
more deeply embedded and hence younger. Similar colour differences are
also seen for sources with \htwoo\ and \methanol\ masers. However
intercomparing the distribution of these colours for sources with each
type maser shows them to be indistinguishable. For the sources in this
study at least, there is no evidence that one or other of the types of
maser in particular traces younger (or older) sources.

The velocity range of the maser emission suggests that the water maser
sources may be divided into two groups, one of which shows narrow
velocity spread of less than $~4-5$ km~\s, while the other shows a
much wider one, with sources which have \htwoo\ maser velocities
spread over up to $\sim65$ km~\s. This division may well reflect the
suggestion by Torrelles et al.  (\citealp{torrelles97},
\citealp{torrelles98}) that \htwoo\ masers can trace both molecular
outflows and accretion disks perpendicular to the molecular outflow.

Comparing the OH and methanol maser peak fluxes, the majority of the
sources are either methanol-favoured or OH-favoured with over 40\% of
the sources in each of these categories.  Both the detection rate and
velocity coverage of the maser emission suggests a close association
between OH and \methanol\ masers, which has also been reported by
other authors (e.g. Caswell \citealp{caswell98}, and references
therein; Szymczak \& Kus \citealp{Szymczak00}; Szymczak \& G\'{e}rard
\citealp{szymczak04}).  However understanding the exact origin of the
OH masers and the detailed connection between the OH and \methanol\
masers requires higher spatial resolution follow-up studies such as
that of IRAS~20126+4104 by Edris et al. (2005).  { High resolution
  observations of some molecular outflow sources have shown that
  towards HMPOs the OH masers often arise from a circumstellar disc
  around the central source (Hutawarakorn, \& Cohen
  \citealp{hutawarakorn99}; Hutawarakorn et al.
  \citealp{hutawarakorn02}; Hutawarakorn et al.
  \citealp{hutawarakorn03}; Edris et al. 2005). } Such high resolution
follow-up of the sources in this survey may therefore identify
additional HMPOs with circumstellar disks and making use of the Zeeman
sensitivity of the OH emission, measure the magnetic field within
them.

\begin{acknowledgement}
  We would like to thank the staff at both \nancay\ and Green Bank for
  their invaluable help in obtaining the data presented here. We would
  also like to thank Malcolm Gray for helpful discussions. K.A.E and
  G.A.F would like to dedicate this paper to the memory of R.J.~Cohen
  who passed away on 1 November 2006.
\end{acknowledgement}




\Online

\onecolumn
\begin{center} {\large \it The figures above are available as a
    separate file from 
    http://www.mananchester.ac.uk/jodrellbank/$\sim$gaf/Papers/6280\_onlinefigs.pdf}
\end{center}


\begin{figure*}
  \caption{The 9-point maps of the maser intensity integrated over the
    velocity range of each maser component detected by the GBT
    observations. The axes show RA and Dec offset in arcminutes from
    the IRAS position. The contours levels run from 10\% to 90 \% (in
    20\% steps) of the peak flux given in Table~\ref{tab:OHmaserRes}.}
  \label{fig:GBT-map}
\end{figure*}

\begin{center}
\begin{figure*}
 \caption{Same as Figure~\ref{fig:GBT-map} but for OH maser sources offset from
 the IRAS position by $>$ 2$^{'}$.}
  \label{fig:GBT-map-offset}
  \end{figure*}
\end{center}

\begin{center}
\begin{figure*}
 \caption{The spectra of each OH maser line detected by GBT and \nancay\
observations. The sources names are given at the top of each group
of spectra. The intensities are in Jy.}
   \label{fig:spectra}
\end{figure*}
\end{center}

 \begin{center}
\begin{figure*}
  \caption{Same as Figure~\ref{fig:spectra} but for sources with
maser emission offset from the IRAS position by $>$ 2$^{'}$.}
  \label{fig:spectra-offset}
\end{figure*}
\end{center}
\clearpage

\appendix
\section{Notes on individual sources where OH masers were detected}
\label{sec:notes on individual sources}
\label{app:sources}

\textit{\textbf{IRAS~05137+3919}}. This source also shows water
maser emission (P91). This IRAS source is one of the sources
studied in detail at high angular resolution by Molinari et al.
\cite{molinari02} at millimeter and centimeter wavelengths in both
continuum and spectral lines. There is a core which shows radio
continuum emission detected at 3.6 cm wavelength (VLA1) and mm
continuum emission detected at 3.4 mm wavelength (Figure 1 of
Molinari et al. \citealp{molinari02}). At 2.2$\mu$m the IRAS
source is resolved in to a cluster of objects (Molinari et al.
\citealp{molinari02}).

\textit{\textbf{IRAS~05274+3345}} (AFGL5142). This source is also
associated with \htwoo\ (Verdes-Montenegro et al.
\citealp{verdes89}; P91) and \methanol\ (SHK2000) masers. OH maser
emission was previously reported towards this source by Braz et
al. \cite{braz90}. The OH lines detected towards this source,
namely 1665, 1667 and 1612 MHz, have varied since the Braz et al.
observations. While the 1665 MHz line has strengthened, the other
two lines have weakened.

\textit{\textbf{IRAS~05358+3543}} (G173.481+2.445, S231, S233IR).
This source has been mapped at high angular resolution,
1.5$^{''}$, with the VLA (see Figure 27 of Argon et al.
\citealp{argon00}). The \nancay$-$GBT observations detected maser
components in the two main lines, 1665 and 1667 MHz, in both the
left and right circular polarisations, while only four components
were reported by Argon et al. in the (LHC) 1665 MHz line. Two
components of these four coincide very well with the IRAS source
and the others are just $\sim$ 0.2$^{''}$ south. Szymczak et al.
\cite{szymczak00b} failed to detect maser emission at the 4765 MHz
OH line towards this source. The IRAS source is also associated
with \htwoo\ (S02) and a strong \methanol\ masers (Galt
\citealp{galt04} and references therein). No radio continuum
detection has been reported towards IRAS~05358+3543 in S02. On the
other hand recent sub$-$millimeter continuum maps revealed a total
of four sources (Williams et al.  \citealp{williams04}; Minier et
al. \citealp{Minier05}). One of these sources is coincident with
the IRAS source and OH maser and $\sim$ 40$^{''}$ north$-$east
another submm source harbours the \methanol\ maser, exhibits
mid$-$infrared emission and coincides with one of the three
1.2$-$mm continuum sources detected by Beuther et al.
\cite{beuther02b} (see figure 4 of Minier et al.
\citealp{Minier05}). High angular resolution observations with
Plateau de Bure Interferometer by Beuther et al. \cite{beuther02d}
reveal that this region contains at least three molecular
outflows.

\textit{\textbf{IRAS~05382+3547}}. The OH maser towards this
source was discovered by Szymczak \& Kus \cite{Szymczak00}, but
no position for it was measured. GBT observations show that the
1665$-$ MHz OH maser components detected by Szymczak \& Kus
centred at velocities -27 and -22 km \s are offset from the IRAS
position by $\sim$ 3$^{'}$. GBT observations found new components
centred at velocities -20 and -24.7 km \s\ offset $\sim$ 30$^{''}$
north$-$west of the IRAS source. This source is also associated
with a \methanol\ maser (SHK2000) with velocity centred at 24.1 km
\s close to one of the OH maser components.

\textit{\textbf{IRAS~06056+2131}}. Cohen et al. \cite{cohen88}
detected OH maser emission at 1665 MHz only. The \nancay$-$GBT
observations found emission at 1667 and 1720 MHz as well.

\textit{\textbf{IRAS~17527$-$2439}}. This source is also
coincident with a \htwoo\ maser (P91).

\textit{\textbf{IRAS~18018$-$2426}} (G6.049$-$1.447, M8E).
This OH maser was first detected by Cohen et al. \cite{cohen88}
and was mapped by Argon et al. \cite{argon00}. These previous
observations detected emission at 1665 MHz only, while
\nancay$-$GBT detected maser emission in the 1667 MHz line as
well. The OH maser emission is variable towards this source (Cohen
et al. \citealp{cohen88}). The water maser was detected towards
this source by Lada et al. \cite{lada76} but P91 failed to detect
any water maser emission. This source is known to be associated
with a compact HII region (Simon et al. \citealp{Simon84}).
Molinari et al. \cite{molinari02} detected 6.75 and 4.9 mJy
continuum radio emission at 2 and 6 cm wavelengths respectively.
The maser towards this source is not polarized which is unusual
(e.g. Cohen \citealp{cohen89}). The disappearance of \htwoo\
masers and the association of OH maser with compact HII region
suggest that this may be in a late stage of star formation where
the UCHII region has expanded and only the OH maser emission is
surviving. No \methanol\ maser towards this source has been
reported.

\textit{\textbf{IRAS~18024$-$2119}}. The newly detected OH main
line maser emission towards this source has a relatively wide
velocity range, covering $\sim40$ km \s. An \htwoo\ maser was also
detected by P91. A relatively strong \methanol\ maser of 100 Jy
was recently detected by Galt \cite{galt04}. No molecular outflow
was detected towards this source by Zhang et al. \cite{zhang05}.

\textit{\textbf{IRAS~18048$-$2019}}. This source shows very weak
OH maser emission in both main lines. Water (P91) and \methanol\
(Schutte et al. \citealp{Schutte93}) masers are also associated
with this source.

\textit{\textbf{IRAS~18089$-$1732}}. This source is common in the
S02 sample and the \textit{High} sub$-$sample of M96. The OH maser
emission was detected by Cohen et al. \cite{cohen88} in the 1665
MHz line only. The Argon et al. \cite{argon00} VLA map shows
several 1665$-$MHz OH maser components in three different
positions. We detected maser emission at 1667 MHz as well. This
source is also associated with \htwoo\ (P91) and \methanol\
(SHK2000) masers and is associated with very weak 3.6 cm continuum
emission, 0.9 mJy (S02). Recent high angular resolution
submillimeter observations in various spectral lines by Beuther et
al. \cite{beuther05} detect a massive rotating structure
perpendicular to an emanating outflow which is likely associated
with the central accretion disk.

\textit{\textbf{IRAS~18090$-$1832}}. This source also shows a
relatively strong, 77 Jy, \methanol\ maser (SHK2000) and weak OH
maser emission in the two main lines.

\textit{\textbf{IRAS~18102$-$1800}}. This source is also associated
with radio continuum emission, 44 mJy, at 3.6 cm wavelength (S02).

\textit{\textbf{IRAS~18144$-$1723}}. This is a relatively strong
OH maser source in the main lines. There is a significant gap, 13
km~\s\ between the central velocities of strongest components in
the 1665 and 1667 MHz lines. The IRAS source is also associated
with \methanol\ (SHK2000) and \htwoo\ (P91) masers. Radio
continuum emission was also detected towards this source at 2 and
6 cm wavelengths (Molinari et al. \citealp{molinari98}).

\textit{\textbf{IRAS~18182$-$1433}}. The \htwoo\ and \methanol\
masers associated with this source were mapped in high angular
resolution, $\sim$ 1$^{''}$ by Beuther et al. \cite{beuther02}.
Only one \methanol\ maser component was detected, while several
\htwoo\ maser components were detected. The \htwoo\ and \methanol\
positions are coincident with the 1.2~mm continuum emission
detected by Beuther et al. \cite{beuther02b}. Extremely weak radio
continuum emission has been detected by Foster \& Caswell
\cite{Foster00} (0.3~mJy at 3.5~cm) and S02 ($<$1 mJy at 3.6~cm).
An outflow was detected towards this
source by Beuther et al. \cite{beuther02c}. 

\textit{\textbf{IRAS~18236$-$1205}}. This IRAS source is also
associated with \methanol\ (SHK2000) and \htwoo\ (P91) masers.

\textit{\textbf{IRAS~18264$-$1152}}. This source shows maser emission
at the 6.7 GHz \methanol\ and 22 GHz \htwoo\ maser lines (SHK2000 and
P91 respectively). No radio continuum emission, $<1$ mJy, is detected
at 3.6 cm wavelength (S02). An outflow was detected towards this
source by Beuther et al.  \cite{beuther02c}.

\textit{\textbf{IRAS~18278$-$1009}}. This source is also
associated with a \methanol\ maser (SHK2000). No radio continuum
emission was detected at 2 and 6 cm wavelengths by Molinari et al.
\cite{molinari98}.

\textit{\textbf{IRAS~18290$-$0924}}. This source is also associated
with \htwoo\ and \methanol\ masers (S02). The mm observations (Beuther
et al. \citealp{beuther02b}) towards this source show two peaks
separated by 12$^{''}$ with the masers coincident with one of them.
This source is also associated with radio continuum emission at 3.6 cm
detected by S02.

\textit{\textbf{IRAS~18310$-$0825}}. This source is the only
source in our sample associated only with the 1667$-$MHz of OH
maser lines. It is also associated with a \methanol\ maser
detected by SHK2000 and mapped by Beuther et al. \cite{beuther02}.
The methanol maser is offset from the IRAS source and coincident
with one of two mm and cm peaks (see Figure 1 of Beuther et al.
\citealp{beuther02}).

\textit{\textbf{IRAS~18316$-$0602}}. This source is also
associated with very strong \htwoo\  (725.83 Jy, P91), and
\methanol\ (178 Jy, SHK2000) masers. This source is known to be
associated with an UCHII region (Jenness et al.
\citealp{Jenness95}) and molecular outflow (Wu et al.
\citealp{wu04}). The OH maser emission is detected in the main
lines while the satellite lines show conjugate behaviour with
thermal emission at 1612 MHz and absorption at 1720 MHz.

\textit{\textbf{IRAS~18345$-$0641}}. Towards this star-forming
region only the 1612 MHz OH maser line was detected.  The IRAS
source is associated with strongly, variable \methanol\ masers
(SHK2000) which coincide with a mm continuum peak (Beuther et al.
\citealp{beuther02}). Also free$-$free emission at 3.6 cm (S02)
and an outflow have been detected towards this IRAS source
(Beuther et al. \citep{beuther02c}).

\textit{\textbf{IRAS~18360$-$0537}}. In addition to the OH main line
masers detected here, this source is also associated with relative
strong \htwoo\ maser emission, 92.73 Jy (P91).

\textit{\textbf{IRAS~18385$-$0512}}. This source is also
associated with a relatively strong \htwoo\ maser emission, 200 Jy
(S02). Several \htwoo\ maser components coincide with a mm
continuum peak (Beuther et al. \citealp{beuther02}). Radio
continuum emission of 29 mJy was measured by S02 at 3.6 cm.

\textit{\textbf{IRAS~18440$-$0148}}. This source is also
associated with a \methanol\ maser (SHK2000; Walsh et al.
\citealp{walsh98}). The \methanol\ maser components coincident
with a 1.2 mm, 3.6 cm peak and mid$-$infrared source (Beuther et
al. \citealp{beuther02}; S02). A tentative detection of \htwoo\
maser emission is reported by S02.

\textit{\textbf{IRAS~18454$-$0158}}. This is one of the sources which
has an OH maser but does not have any other known masers.  S02
observations did not detect \htwoo\ or \methanol\ maser emission. This
source is associated with 1.2 mm continuum emission as well as radio
continuum emission (Beuther et al \citealp{beuther02b} and S02
respectively).

\textit{\textbf{IRAS~18463+0052}}. Only 1612~MHz line OH masers
are detected towards this source with a line profile with peaks at
67 and 92 km \s, suggesting this source is an OH/IR star rather
than a star-forming region.  There are no \htwoo\ or \methanol\
masers associated with the IRAS source.

\textit{\textbf{IRAS~18488+0000}}. This is one of the common
sources in the S02 sample and the \textit{High} sub$-$sample of
M96. This source is associated with a variable \methanol\ maser
(SHK2000, and references therein). Although not detected by P91,
\htwoo\ maser emission has been detected by S02. The water maser
is coincident within few arcsecond with a mid$-$infrared source but
offset from a millimetre continuum source (Beuther et al. \citealp{beuther02}).
Relatively strong radio continuum emission, 194 mJy, has been
detected towards this source (S02). Our \nancay$-$GBT observations
detect OH maser emission in the two main lines which is coincident
(within $\sim$ 30$^{''}$) with the 1.2 mm emission source rather
than other tracers.

\textit{\textbf{IRAS~18507+0121}}. This source is roughly 11$^{'}$
north from G34.257+0.154 (or G34.3+0.2) which shows strong OH
maser emission in the main lines as well as 1720 MHz satellite
line (Argon et al. (\citealp{argon00}). G34.257+0.154 also has
\htwoo\ masers and is classified as an HII region (Benson \&
Johnston \citealp{benson84}; Genzel \& Downes \citealp{genzel77}).
The \nancay$-$GBT observations detect new (relatively weaker) OH
maser emission associated with the IRAS source in the main lines.
The IRAS source is also associated with \htwoo\ (P91) and
\methanol\ (SHK2000) masers. IRAS~18507+0121 region was studied in
detail by Shepherd et al \cite{Shepherd04} at several millimeter
and near$-$infrared (NIR) wavelengths. Shepherd et al detected two
compact molecular cores separated by $\sim$ 40$^{''}$ north-south.
The northern molecular core contains a newly discovered, deeply
embedded, B2 protostar surrounded by several hundred solar masses
of warm gas and dust, G34.4+0.23 MM. Based on the presence of warm
dust emission and the lack of detection at NIR wavelengths,
Shepherd et al suggest that G34.4+0.23 MM may represent the
relatively rare discovery of a massive protostar (analogous to a
low$-$mass "Class~0" protostar).  The southern molecular core is
associated with an NIR cluster of young stars and an UCHII region,
G34.4+0.23 (detected by Miralles, Rodr\'{\i}guez \& Scalise
\citealp{Miralles94}), with a central B0.5 star.  Shepherd et al
to suggest an upper limit on the age of the IRAS~18507+0121
star--forming region of 3 Myr. This IRAS source is not associated
with molecular outflow (Zhang et al. \citealp{zhang05}).

\textit{\textbf{IRAS~18527+0301}}. This source was searched for OH
masers by Szymczak \& Kus \cite{Szymczak00} to a rms noise level
of 0.2 Jy but none were detected. The present observations, with
better sensitivity (0.02 Jy), detected weak emission in the two
main lines. Methanol maser emission was detected by SHK2000 in the
same OH velocity range. No radio continuum emission at 6~cm was
detected by Molinari et al. \cite{molinari98}.


\textit{\textbf{IRAS~18553+0414}}. This source is also associated with
\htwoo\ maser emission (S02) which is coincidence with a millimeter
continuum source (Beuther et al. \citealp{beuther02b}).

\textit{\textbf{IRAS~18566+0408}}. This common source in the S02
sample and the \textit{High} sub$-$sample of M96 is also
associated with \methanol\ maser emission (S02; SHK2000). The
\htwoo\ maser emission was newly detected by S02, while not
detected by P91. The \htwoo\ masers are in better agreement with a
millimeter continuum (Beuther et al. \citealp{beuther02}). The
newly detected OH maser emission also seemingly coincides with the
mm source mapped by Beuther et al. \cite{beuther02b}. An outflow
was detected by Beuther et al. \cite{beuther02c} and S02 place an
upper limit of 1 mJy on the 3.6 cm radio continuum flux from any
source in this region.

\textit{\textbf{IRAS~19035+0641}} (G40.622$-$0.137). This source
is one of OH maser sources mapped with the VLA by Argon et al.
\cite{argon00}. Several components, at different velocities in the
range of 25 to 36 km~\s, were detected in the two main lines and
spread over $\sim$ 1 arcsec (figure 14 of Argon et al.
\citealp{argon00}). Our \nancay$-$GBT observations show similar
emission although the flux density of the components has varied
since the Argon et al.  observations.  The flux density of the
1667~MHz RHC compoment centered at velocity 27.39 km~\s\ has
rocketed up from 0.44 to 22.3 Jy. The IRAS source is also
associated with \htwoo\ and \methanol\ masers (S02, SHK2000 and
references therein). The \htwoo\ and \methanol\ masers are
coincident with a mid$-$infrared and 1.2mm continuum emission
sources (see figure 1 of Beuther et al. \citealp{beuther02}). The
UCHII region, detected at 6$-$cm by Hughes \& Macleod
\cite{Hughes93} and 3.6$-$cm by S02, shows weak emission, of 3.8
and 4 mJy respectively. An outflow was detected towards this
source by Beuther et al.  \cite{beuther02c}.

\textit{\textbf{IRAS~19092+0841}}. This source, also associated with
\htwoo\ (P91) and \methanol\ (SHK2000) masers, is one of the newly
detected of OH masers. Radio continuum emission of 2.74 and 1.04 mJy
have been detected by Molinari et al. \cite{molinari98} at the 2$-$
and 6$-$cm wavelengths respectively.

\textit{\textbf{IRAS~19118+0945}}. This is one of the sources with only OH
masers. No \htwoo\ or \methanol\ masers have been detected
towards this source (P91; SHK2000).

\textit{\textbf{IRAS~19217+1651}}. This source is also associated with
\methanol\ masers and recently detected \htwoo\ masers as well as radio
continuum emission (S02). One mm source was detected by Beuther et al.
\cite{beuther02b} $\sim$ 5$^{''}$ north of the IRAS source and consistent with
the \htwoo\ and \methanol\ masers and a mid$-$infrared source. The radio
continuum source is $\sim$ 5$^{''}$ west of the mm source. The detected OH
maser emission seems not consistent with any of the previous sources, being
located $\sim$ 1$^{'}$ to the south. Beuther et al. \cite{beuther04} studied
this region in detail with high spatial resolution using Plateau de Bure
Interferometer in the CO J=2$-$1 and SiO J=2$-$1 transitions.  They conclude
that the high$-$mass region IRAS~19217+1651 exhibits a bipolar outflow and the
region is dominated by the central driving source.

\begin{figure}
\resizebox{\hsize}{!}{\includegraphics[angle=-90]{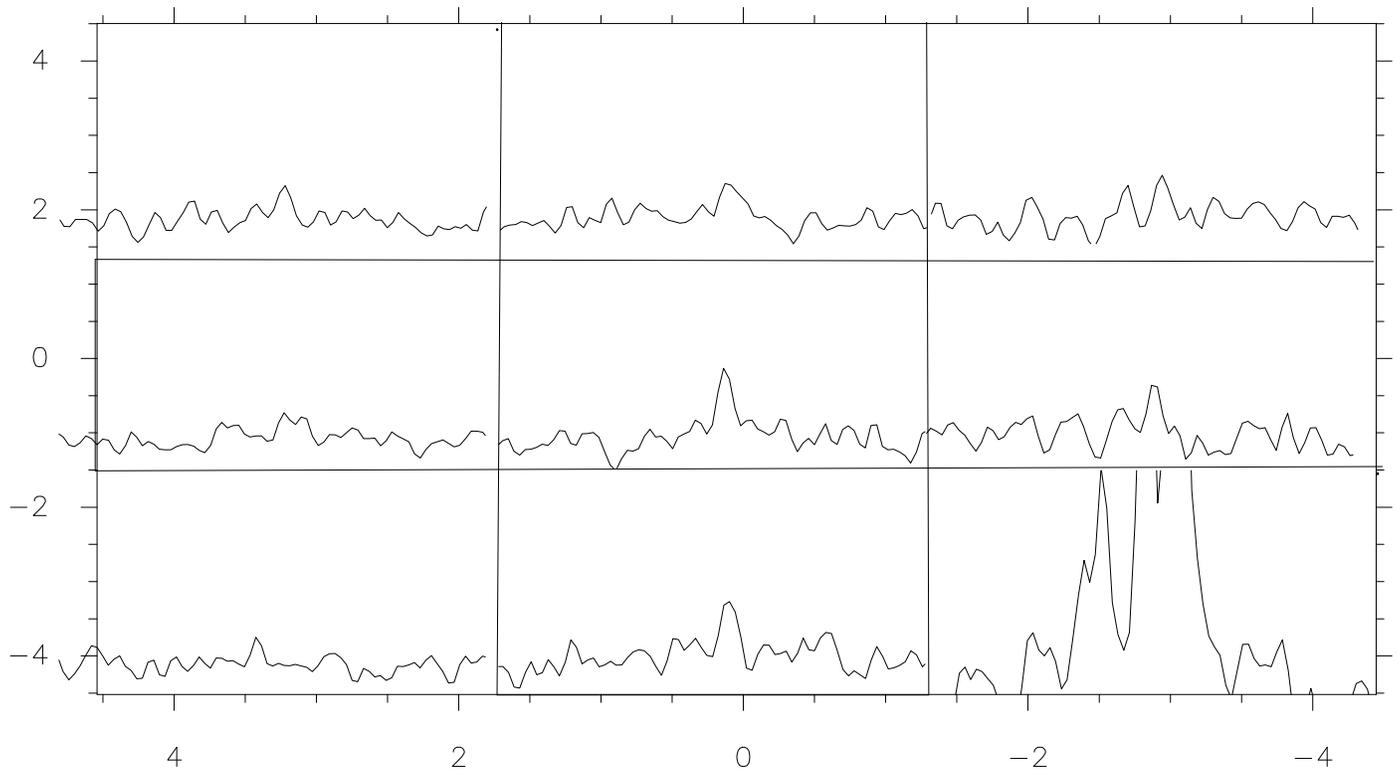}}
\caption{The spectra from the IRAS19220+1432 map in the 1720 MHz
OH maser. The OH maser associated with the IRAS position (middle
panel) is contaminated by the strong OH maser source W51 which
appears in the lower right hand side panel of the map. The axes
show offsets in arcminutes from the IRAS position.}
 \label{19220-spec-map}
\end{figure}

\textit{\textbf{IRAS~19220+1432}}. The OH maser towards this
source is contaminated by the nearby OH maser source W51.  However
the GBT map (Figure~\ref{fig:GBT-map}) shows that there is a maser
emission in the 1720$-$MHz (RHC) line associated with the IRAS
position. This is better illustrated in
Figure~\ref{19220-spec-map} where the spectra from the map are
plotted. The contamination of W51 maser is clear in the lower
right corner of the map, but a component centred on the IRAS
source is also visible. There is no other maser types associated
with this source (S02). An flux density of 11~mJy in the radio
continuum was detected by S02 at 3.6$-$cm wavelength.

\textit{\textbf{IRAS~19374+2352}}. This source is also associated with
a \htwoo\ maser (P91), but no \methanol\ maser emission was detected
by SHK2000. Free$-$free emission at 64.4 mJy was detected by Molinari
et al. \cite{molinari98}.

\textit{\textbf{IRAS~19388+2357}}. This source was searched by
Szymczak \& Kus \cite{Szymczak00} for OH maser but no emission
was found. The observations here detect OH maser emission in the
1665$-$MHz line spread over velocity range 34 to 39 km~\s. On the
other hand, \methanol\ maser emission, although detected by
Schutte et al.  \cite{Schutte93} in 1992 and Slysh et al.
\cite{Slysh99} in 1995, was not detected in later observations by
SHK2000 in 1999. This source is also associated with \htwoo\ maser
(P91) and free$-$free emission (Molinari et al.
\citealp{molinari98}).

\textit{\textbf{IRAS~19410+2336}}. This source is also associated
with \htwoo\ and \methanol\ masers (S02; SHK2000). Near the
\htwoo\ maser source, a radio continuum emission of 1 mJy was
detected by S02. Beuther et al. \cite{beuther02b} 1.2$-$mm
continuum observations detect one source consistent with a
mid$-$infrared source. The 1665$-$MHz OH maser emission detected
close to the mid$-$infrared source. An outflow was detected
towards this source by Beuther et al. \cite{beuther02c}.

\textit{\textbf{IRAS~20062+3550}}. This source was detected by
Szymczak \& Kus \cite{Szymczak00} at the velocity of $-$2.4
km~\s\ in the 1665$-$MHz line. The observations reported here
detect a component in the same line over the velocity range $-$1
to 2.2 km~\s. Water and \methanol\ masers have also been detected
towards this source (P91; Slysh et al. \citealp{Slysh99};
SHK2000). No radio emission has been detected at 6 cm towards this
source by Molinari et al. \cite{molinari98}. This source is one of
sources studied in detail by Molinari et al.  \cite{molinari02}
with the Owens Valley Radio Observatory (OVRO) millimeter wave
array. Four distinct cores were identified in the HCO$^{+}$
J=1$-$0. Two of them are also detected in H$^{13}$CO$^{+}$
J=1$-$0. One core also has a 3.4 mm counterpart and is likely the
most massive member of this cluster (Molinari et al.
\citealp{molinari02}).

\textit{\textbf{IRAS~20126+4104}}.  This source has a luminosity
of $10^4$ \Lsun\ and is perhaps the best studied example of a
massive protostar associated with a Keplerian disk and a
jet/outflow system (Cesaroni et al.  \citealp{cesaroni97}, 1999;
Hofner et al. \citealp{hofner99}; Zhang et al.  \citealp{zhang98};
Cohen et al. \citealp{cohen88}; Tofani et al.  \citealp{tofani95};
Moscadelli et al.  \citealp{moscadelli00}). The source is
associated with OH, \htwoo\ and \methanol\ masers.  Observations
of the water masers using the VLA with angular resolution of
$0.1''$ identified three emission regions (Tofani et al.
\citealp{tofani95}).  Moscadelli et al. \citealp{moscadelli00}
resolved two of these into 26 unresolved spots using the VLBA. The
velocity and spatial structure of these spots were well fitted by
a model with the spots arising at the interface between a jet and
the surrounding molecular gas.  Two features of OH masers were
first detected in the 1665$-$MHz line by Cohen et al.
\cite{cohen88}. More recently mapping of the OH and \methanol\
masers at high angular resolution using MERLIN, Edris et al.
\cite{Edris05} showed that OH and methanol masers appear to trace
part of the circumstellar disk around the central source.

\textit{\textbf{IRAS~20188+3928}}. Only the 1720~MHz of four OH
lines is detected towards this source. The source is also
associated with a \htwoo\ maser (P91) and a 2.86 Jy 6~cm radio
continuum source (Molinari et al \cite{molinari98}).

\textit{\textbf{IRAS~20227+4154}}. This source is associated with a \htwoo\
maser (P91) but no \methanol\ maser emission was detected by SHK2000.
Only the 1665~MHz OH maser line has been detected by our \nancay$-$GBT
observations.

\textit{\textbf{IRAS~22198+6336}}. This source is also associated
with a \htwoo\ maser emission (P91) but no \methanol\ maser emission
was detected by SHK2000. No radio emission was detected by
Molinari et al \cite{molinari02} at 6~cm wavelength. The two main
lines of OH maser were detected by our \nancay$-$GBT observations.

\textit{\textbf{IRAS~22272+6358}}. Although not detected in 1993
to a 3$\sigma$ upper limit of about 0.15 Jy (Slysh et al.
\citealp{Slysh94}), OH emission was found in 1999 in both main
lines by Szymczak \& Kus \cite{Szymczak00}. This suggests
considerable variations of the source.  Approximately three years
after Szymczak \& Kus observations, the \nancay$-$GBT observations
show that the OH masers have varied.  At the 1665~MHz (LHC), a new
component was detected centred at 8.46 km~\s\ and the velocity of
a bright component has slightly varied by $\sim$ 1.2 km~\s\ (from
$-$10.9 to $-$12.12 km~\s\ ).
The velocity range of the OH emission is similar to that observed
for the 6.7 GHz methanol maser (SHK2000). No \htwoo\ maser was
detected by P91.

\textit{\textbf{IRAS~23139+5939}}. This sources is very similar to
IRAS 18345$-$0641. Only the 1612 MHz OH maser line has been
detected and the source is also associated with 3.6$-$cm radio
continuum emission (S02). It is also associated with a \htwoo\
maser (S02) which coincides with a mm continuum emission and
mid-infrared sources (Beuther et al. \citealp{beuther02}). An
outflow along the line of sight was detected by Beuther et al.
\cite{beuther02c}.


\begin{thebibliography}{}

\bibitem[1978]{altenhoff78}
  Altenhoff, W.J., Downes, D., Pauls, T., Schraml, J. 1978, A\&AS,
  35, 23

\bibitem[2000]{argon00}
  Argon, A.L., Reid, M.J., Menten, K.M. 2000, ApJS, 129, 159

\bibitem[1996]{Bachiller96}
  Bachiller, R. 1996, ARA\&A, 34, 111

\bibitem[1984]{benson84}
  Benson, J.M. \& Johnston, K.J. 1984, ApJ, 277,181

\bibitem[2002]{beuther02}
  Beuther, H., Walsh, A., Schilke, P., Sridharan, T. K., Menten, K. M.,
  Wyrowski, F. 2002, A\&A, 390, 289

\bibitem[2002c]{beuther02c}
  Beuther, H., Sridharan, T. K., Schilke, P., et al. 2002b, A\&A, 383,
  892

\bibitem[2002b]{beuther02b}
  Beuther, H., Schilke, P., Menten, K.M. et al. 2002c, ApJ, 566,
  945

\bibitem[2002d]{beuther02d}
  Beuther, H., Schilke, P., Gueth, F. et al. 2002d, A\&A, 387,
  931

\bibitem[2004]{beuther04}
  Beuther, H., Schilke, P., Gueth, F. 2004, ApJ, 608, 330

\bibitem[2005]{beuther05}
  Beuther, H., Zhang, Q., Sridharan, T. K., Chen, J 2005, ApJ, 628,
  800

\bibitem[2005]{bourke05}
  Bourke, T.,L., Hyland, A.R., Robinson, G. 2005, ApJ, 625, 883

\bibitem[1990]{braz90}
  Braz, M.A., Lepine, L.R.D., Sivagnanam, P., Le Squeren, A.M. 1990,
  A\&A, 236, 479

\bibitem[1988]{brebner88}
  Brebner, G.C. 1988 \textit{PhD Thesis} University of Manchester

\bibitem[1996]{Bronfman96}
  Bronfman, L., Nyman, L.A., May, J. 1996, A\&AS, 115, 81

\bibitem[1983]{caswell83}
  Caswell, J.L., Haynes, R.F. 1983, Aust.J.Phys., 36, 361

\bibitem[1995]{caswell95}
  Caswell J. L., Vaile R. A., Forster J. R . 1995, MNRAS, 277, 210

\bibitem[1996]{caswell96}
  Caswell, J.L. 1996, MNRAS, 279, 79

\bibitem[1997]{caswell97}
  Caswell, J.L. 1997, MNRAS, 289, 203

\bibitem[1998]{caswell98}
  Caswell, J.L. 1998, MNRAS, 297, 215

\bibitem[1997]{cesaroni97}
  Cesaroni, R., Felli, M., Testi, L., Walmsley, C. M., Olmi, L. 1997,
  A\&A, 325, 725 (C97)

\bibitem[2002]{churchwell02}
  Churchwell, E., 2002, Annu. Rev. Astron. Astrophys. 40, 27

\bibitem[1984]{cohen84}
  Cohen, R. J., Rowland, P.R., Blair, M.M. 1984, MNRAS, 210, 425

\bibitem[1988]{cohen88}
  Cohen, R. J., Baart, E.E., Jonas, J.L. 1988, MNRAS, 231, 205

\bibitem[1989]{cohen89}
  Cohen, R.J. 1989, Rep. Prog. Phys., 52, 881

\bibitem[2003]{cohen03}
  Cohen, R.J., Brebner, G.C., Hutawarakorn, B., Gasiprong, N.,
  IAU General Assembly, 2003 IAUS, 221P, 168C

\bibitem[2002]{cragg02}Cragg, D. M., Sobolev, A. M.,
  Godfrey, P. D.  2002, MNRAS, 331, 521

\bibitem[2005]{Cragg05}
  Cragg, D. M., Sobolev, A. M., Godfrey, P. D. 2005, MNRAS, 360,
  533

\bibitem[2005]{Edris05}
  Edris, K.A., Fuller, G.A., Cohen, R.J., Sandra, E., 2005 A\&A,
  343, 213

\bibitem[1976]{Elitzur76}
  Elitzur, M. 1976, ApJ, 203, 124

\bibitem[1978]{elitzur78}
  Elitzur, M., de Jong, T. 1978, A\&A, 67, 323


\bibitem[2005]{etoka05}
  Etoka, S. Cohen, R.J., \& Gray, M.D. 2005, MNRAS, 360, 1162

\bibitem[2000]{Foster00}
  Foster, J.R., \& Caswell, J.L. 2000, ApJ, 530, 371

\bibitem[2004]{galt04}
  Galt, J 2004, AJ, 127, 3479

\bibitem[1999]{garay99}
  Garay, G., Lizano, S. 1999, PASP, 111, 1049

\bibitem[1977]{genzel77}
  Genzel, R. \& Downes, D. 1977, A\&AS, 30, 145

\bibitem[1992]{gray92} Gray, M.~D., Field, D., \& 
Doel, R.~C.\ 1992, \aap, 262, 555 

\bibitem[1991]{Gregory91}
  Gregory, P.C., Condon, J.J. 1991, ApJS, 75, 1011

\bibitem[1994]{Griffith94}
  Griffith, M.R., Wright, A.E., Burke, B.F., Ekers, R.D. 1994,
  ApJS, 90, 179

\bibitem[1979]{haynes79}
  Haynes, R.F., Caswell, J.L., Simons, L.W.J. 1979, Aust. J. Phys.
  Astrophys. Suppl. 48, 1

\bibitem[1999]{hofner99}
  Hofner, P., Cesaroni, R., Rodr\'{i}guez, L. F., \& Marti J. 1999, AA, 345, 43

\bibitem[1993]{Hughes93}
  Hughes, V.A., \&  Macleod, G.C. 1993, AJ, 105, 1495

\bibitem[1999]{hutawarakorn99}
  Hutawarakorn, B., Cohen, R. J. 1999, MNRAS, 303, 845

\bibitem[2002]{hutawarakorn02}
  Hutawarakorn, B., Cohen, R. J., Brebner, G. C. 2002, MNRAS, 330, 349

\bibitem[2003]{hutawarakorn03}
  Hutawarakorn, B., Cohen, R. J. 2003, MNRAS, 345, 175

\bibitem[1995]{Jenness95}
  Jenness, T., Scott, P.F., Padman, R. 1995, MNRAS, 276, 1024

\bibitem[1976]{lada76}
  Lada, C.J. 1976, ApJS, 32, 603

\bibitem[1996]{Liechti96}
  Liechti, S. \& Wilson, T.L. 1996, A\&A, 314, 615

\bibitem[1999]{Lockett99}
  Lockett, P., Gauthier, E., Elitzur, M. 1999, ApJ, 447, 211

\bibitem[1974]{marsalkova74}
  Marsalkova, P. 1974, Ap\&SS, 27, 3

\bibitem[2001]{minier01}
  Minier, V., Conway, J. E., Booth, R. S. 2001, A\&A, 369, 278

\bibitem[2005]{Minier05}
  Minier, V., Burton, M.G., Hill, T., Pestalozzi, M.R., Purcell,
  C.R., et al 2005, A\&A, 429, 945

\bibitem[1994]{Miralles94}
  Miralles, M.P., Rodr\'{i}guez, L.F., \& Scalise, E. 1994, ApJS,
92, 173

\bibitem[1996]{molinari96}
  Molinari, S., Brand, J., Cesaroni, R., Palla, F. 1996, A\&A, 308, 573

\bibitem[1998]{molinari98}
  Molinari, S., Brand, J., Cesaroni, R., Palla, F., Palumbo, G.G.C. 1998, A\&A, 336,
  339

\bibitem[2002]{molinari02}
  Molinari, S., Testi, L., Rodr\'{i}guez, L., Zhang, Q. 2002, ApJ, 570,
  758

\bibitem[1988]{moore88}
  Moore, T.J.T., Mountain, C.M., Yamashita, T., \& Selby, M.J. 1988,
  MNRAS, 234, 95

\bibitem[2000]{moscadelli00}
  Moscadelli, L., Cesaroni, R., Rioja, M. J. 2000, A\&A, 360, 663 (MCR)

\bibitem[1993]{Palagi93}
  Palagi, F., Cesaroni, R., Comoretto, G., Felli, M., Natale, V.
  1993, A\&AS, 101, 153

\bibitem[1991]{palla91}
  Palla, F., Brand, J., Cesaroni, R., Comoretto, G., Felli, M.,
  1991, A\&A, 246, 249 (P91)


\bibitem[2005]{pest2005} Pestalozzi, M.~R., 
Minier, V., \& Booth, R.~S.\ 2005, \aap, 432, 737 

\bibitem[1987]{richards87}
  Richards, P.J., Little, L.T., Toriseva, M., Heaton, B.D., 1987,
  MNRAS 228, 43

\bibitem[1993]{Schutte93}
  Schutte, A.J., Van der Walt, D.J., Gaylard, M.J., Macleod, G.C.,
  1993, MNRAS, 261, 783

\bibitem[2004]{Shepherd04}
  Shepherd, D.S., N\"{u}rnberger, D.E.A., Bronfman, L. 2004, ApJ,
  602, 850

\bibitem[1987]{shu87}
  Shu, F.H., Adams, F.C., Lizano, S. 1987, ARA\&A, 25, 23

\bibitem[1984]{Simon84}
  Simon, M., Cassar, L., Felli, M., Fischer, J., Massi, M., Sanders, D. 1984,
  ApJ, 278, 170

\bibitem[1994]{Slysh94}
  Slysh, V.I., Dzura, A.M., Val'tts, I.E., G\'erard, E., 1994, A\&AS
106, 87

\bibitem[1997]{Slysh97}
  Slysh, V.I., Dzura, A.M., Val'tts, I.E., G\'erard, E., 1997, A\&AS
124, 85

\bibitem[1999]{Slysh99}
  Slysh, V.I., Val'tts, I.E., Kalenskii, S.V., et al. 1999, A\&AS 134,
  115

\bibitem[2002]{sridharan02}
  Sridharan, T. K., Beuther, H., Schilke, P., Menten, K. M.,
  Wyrowski, F. 2002, ApJ, 566, 931

\bibitem[2004]{szymczak04}
  Szymczak, M., \& G\'erard, E. 2004, A\&A, 414, 235

\bibitem[2000]{SHK2000}
  Szymczak, M., Hrynek, G., \& Kus, A.J. 2000, A\&AS, 143, 269  (SHK2000)

\bibitem[2000b]{szymczak00b}
  Szymczak, M., Kus, A.J. \& Hrynek, G. 2000b, MNRAS, 312, 211 

\bibitem[2000c]{Szymczak00}
  Szymczak, M., Kus, A.J. 2000, A\&AS, 147, 181

\bibitem[1995]{tofani95}
  Tofani, G., Felli, M., Taylor, G. B., Hunter, T. R. 1995, A\&AS 112, 299

\bibitem[1997]{torrelles97}
  Torrelles J.M., G\'{o}mez J.F., Rodr\'{i}guez L.F., et al., 1997, ApJ
489, 744

\bibitem[1998]{torrelles98}
  Torrelles J.M., G\'{o}mez J.F., Rodr\'{i}guez L.F., et al., 1998, ApJ 505,
756

\bibitem[1995]{Valtts95}
  Val'tts I.E., Dzyura, A.M., Kalenskii, S.V., Slysh, V.I., Bus, R., Vinnberg, A.
  1995, ARep, 39, 18

\bibitem[1989]{verdes89}
  Verdes-Montenegro, L., Torelles, J.M., Rodriguez, L.F., Anglada, G.,
  Lopez, R., Estalella, R., Cant\'{o}, J., Ho, P.T.P., 1989, ApJ, 346, 193

\bibitem[1998]{walsh98}
  Walsh, A.J., Burton, M.G., Hyland, A.R., \& Robinson, G. 1998,
  MNRAS, 301, 640

\bibitem[2004]{williams04}
  Williams, S.J., Fuller, G.A., and Sridharan, T.K. 2004, A\&A, 417, 115

\bibitem[1989]{wood89}
  Wood, D.O.S., \& Churchwell, E. 1989, ApJS., 83,119 (WC89)


\bibitem[1994]{Wright94}
  Wright, A.E., Griffith, M.R., Burke, B.F., Ekers, R.D. 1994, ApJS, 91,
  111



\bibitem[2004]{wu04}
  Wu, Y., Wei, Y., Zhao, M., Shi, Y., Yu, W., Qin, S., Huang, M.
  2004, A\&A, 426, 503

\bibitem[1998]{zhang98}
  Zhang, Q., Hunter, T. R., Sridharan, T. K. 1998, ApJ, 505, 151

\bibitem[2005]{zhang05}
  Zhang, Q., Hunter, T. R., Brand, J., et al. 2005, ApJ, 625, 864

\end{thebibliography}
\end{document}